\documentclass[aps,prl,floatfix,twocolumn,showpacs,preprintnumbers,amsmath,amssymb,superscriptaddress]{revtex4-1}
\usepackage{graphicx}
\usepackage{bm}
\usepackage{epstopdf}

\begin{document}
\title{Tunable Magnetic Textures: From Majorana Bound States to Braiding}
\author{Alex Matos-Abiague}
\affiliation{Department of Physics, University at Buffalo, State University of New York, Buffalo, NY 14260, USA}
\author{Javad Shabani}
\affiliation{Department of Physics, City College of the City University of New York, New York, New York 10031, USA}
\author{Andrew D. Kent}
\affiliation{Department of Physics, New York University, New York, New York 10003, USA}
\author{Geoffrey L. Fatin}
\affiliation{Department of Physics, University at Buffalo, State University of New York, Buffalo, NY 14260, USA}
\affiliation{Department of Physics, Cornell University, Ithaca, New York 14853, USA}
\author{Benedikt Scharf}
\affiliation{Department of Physics, University at Buffalo, State University of New York, Buffalo, NY 14260, USA}
\affiliation{Institute for Theoretical Physics and Astrophysics, University of W\"{u}rzburg, Am Hubland, 97074 W\"{u}rzburg, Germany}
\author{Igor \v{Z}uti\'c}
\affiliation{Department of Physics, University at Buffalo, State University of New York, Buffalo, NY 14260, USA}

\date{\today}
 
\begin{abstract}
A versatile control of magnetic systems, widely used to store information, can also enable
manipulating Majorana bounds states (MBS) and implementing fault-tolerant quantum information processing. The proposed platform relies on the proximity-induced superconductivity in
a two-dimensional electron gas placed next to an array of magnetic tunnel junctions (MTJs).  A change in the magnetization configuration in the MTJ array creates tunable magnetic textures thereby removing several typical requirements for MBS: strong spin-orbit coupling, applied magnetic field, and confinement by one-dimensional structures which complicates demonstrating non-Abelian statistics through braiding. Recent advances in fabricating two-dimensional epitaxial superconductor/semiconductor heterostructures and designing tunable magnetic textures support the feasibility of this novel platform for MBS.

\end{abstract}

\maketitle

\vspace{-.2cm}
\subsection{1. Introduction}
\vspace{-.2cm}
Emergent and topologically-nontrivial quasiparticles in superconductors, such as Majorana bound states (MBS), are neither Fermions, nor Bosons. Instead, exchanging MBS yields a non-commutative phase, a sign of non-Abelian statistics and non-local degrees of freedom to implement fault-tolerant quantum information processing \cite{Nayak2008:RMP,Kitaev2003:AP}. While superconductors combine desirable quantum coherence and an energy gap, a spinless $p$-wave electron pairing required for MBS remains elusive \cite{Ivanov2001:PRL,Mackenzie2003:RMP,Sengupta2001:PRB,Wu2010:PRB}. Even among potential $p$-wave candidates their pairing symmetry may have an alternative explanation \cite{Zutic2005:PRL}. Rather than seeking an inherent 
$p$-wave superconductor, a common approach to realize MBS is to use conventional $s$-wave superconductors and engineer their heterostructures with a suitable symmetry of a proximity-induced superconductivity \cite{Fu2008:PRL,Lutchyn2010:PRL,Sau2010:PRL,Oreg2010:PRL}. To satisfy the particle-antiparticle symmetry of MBS and realize effective spinless pairing, most of the proposals combine an applied magnetic field and the native spin-orbit coupling (SOC) in semiconductor nanowires which are also important to localize MBS \cite{Alicea2012:RPP,Leijnse2012:SST}. However, such one-dimensional (1D) structures are inherently limited. In 1D the evidence of MBS detection is indirect, typically relying on a zero-bias conductance peak \cite{Mourik2012:S,Deng2012:NL,Rokhinson2012:NP,Das2012:NP,Finck2013:PRL,NadjPerge2014:S,Pawlak2016:NPJQI,Deng2016:S}, rather than probing directly their non-Abelian statistics. The existing 1D geometries also pose additional obstacles to realize braiding and fusing of MBS, the key elements for topological quantum computing. As an alternative, the use of complex quantum wire networks has been proposed \cite{Alicea2011:NP,Aasen2016:PRX}.

Starting from 1D systems, the realization of MBS in 2D requires a pathway beyond the mere extension of the system dimension. For example, in semiconductor wires with SOC the MBS do not survive the transition to 2D but rather evolve into edge states when the wire width increases \cite{Potter2010:PRL,Sedlmayr2015:PRB}. In the absence of SOC MBS can still emerge in superconducting nanowires under the effects of a helical magnetic texture (see discussion below) \cite{Klinovaja2012:PRL,Kjaergaard2012:PRB,Pientka2013:PRB,NadjPerge2014:S,Pawlak2016:NPJQI}. In such a case the MBS may survive the transition to 2D but spread entirely along opposite edges, precluding braiding.

In this work we explore an alternative path to realize and control MBS based on tunable magnetic textures. They not only obviate the need for an applied magnetic field, but at the same time induce synthetic SOC and enable localizing and moving MBS in a 2D electron gas (2DEG), a natural setting to implement the non-Abelian braiding statistics under exchange which would provide both an ultimate proof for the MBS existence and the key element for topological quantum computing.

Since proposing tunable magnetic textures to control MBS in 2D \cite{Fatin2016:PRL}, several recent experimental 
advances~\cite{Shabani2016:PRB,Suominen2017:P} have further supported the feasibility of such a platform 
as well as put forth different materials implementations which we consider here. 
Desirable semiconductors for this platform could be realized without magnetic doping considered in Ref.~\onlinecite{Fatin2016:PRL}.
In addition to discussing the main
experimental challenges, our transparent description of the role of magnetic textures as well as their distinguishing 
features in 2D,  provides a guidance for  employing other 2D systems to realize MBS, not limited to conventional semiconductors.

\vspace{-.2cm}
\subsection{2. Tunable Magnetic Textures}
\vspace{-.2cm}
Most of previous investigations have been focused on the use of electrostatics for manipulating MBS \cite{Alicea2011:NP,Kim2015:PRB,Sau2011:PRB,Clarke2011:PRB,Halperin2012:PRB,Klinovaja2013:PRX,Aasen2016:PRX}. However, the use of magnetic textures may provide some advantages. As shown below, magnetic textures can generate both Zeeman-like fields and synthetic SOC and, at the same time, induce particle confinement, overcoming the need for complex networks of physical wires. Within this scheme, the manipulation of MBS is realized by properly tuning 
the magnetic textures, without the need of additional contacts and their corresponding risk of particle poisoning \cite{Aasen2016:PRX}. 

For illustration we consider a 2DEG with chemical potential $\mu$, in the presence of a magnetic texture $\mathbf{B}(\mathbf{r})$. The system can be described by the Hamiltonian,
\begin{equation}\label{hamiltonian}
H = \left(\frac{p^2}{2m^\ast} -\mu \right) + \mathbf{J}(\mathbf{r})\cdot\boldsymbol{\sigma}\;,
\end{equation}
where $m^\ast$ and $\mathbf{p}$ are, respectively, the effective mass and kinetic momentum (for simplicity, we neglect orbital effects), and $\boldsymbol{\sigma}$ is the vector of Pauli matrices. The last term in the Hamiltonian corresponds to the Zeeman interaction, $\mathbf{J}(\mathbf{r})=g_{\rm eff} \mu_{\rm B} \mathbf{B}(\mathbf{r})/2$, where
$g_{\rm eff}$ is the effective $g$-factor and $\mu_{\rm B}$ denotes the Bohr magneton.

\begin{figure}[t]
	\centering
	\includegraphics*[width=7cm]{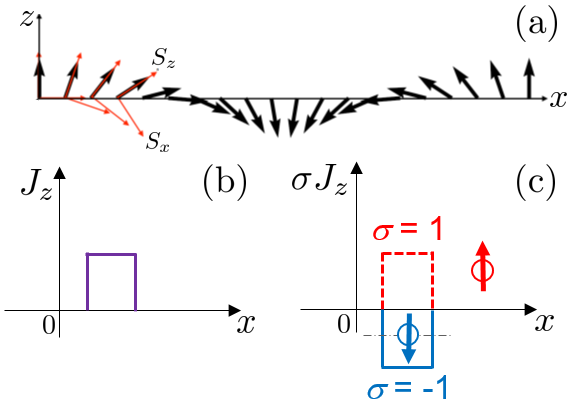}
	\caption{(a) Local rotations of the spin axes in a magnetic texture. (b) A piecewise constant strength of a magnetic texture. (c) Corresponding spin-down confinement induced by an inhomogeneous magnetic texture.
	}\label{fig:texture}
\end{figure}

The Zeeman interaction can be diagonalized by performing local spin rotations aligning the spin quantization axis to the local magnetic field direction [see Fig.~\ref{fig:texture}(a)].
A similar procedure has been known for over 40 years and used in a variety of physical systems 
\cite{Korenman1977:PRB,Tatara1997:PRL,Bruno2004:PRL,Jia2010:PRB}, while
more recently it was employed in the studies of 1D systems \cite{Braunecker2010:PRB,Klinovaja2013:PRL,Vazifeh2013:PRL} where it could be suitable for realizing MBS. In the rotated frame, the Hamiltonian reads as
\begin{equation}\label{h-rotated}
H'=\frac{\left(\mathbf{p}-e\boldsymbol{ \mathbb{A}}(\mathbf{r})\right)^2}{2m^*}-\mu + J(\mathbf{r}) \sigma_z\;,
\end{equation}
where $J(\mathbf{r})=|\mathbf{J}(\mathbf{r})|$, and $\boldsymbol{\mathbb{A}}(\mathbf{r})$ is the resulting non-Abelian field that yields the synthetic SOC,
\begin{equation}\label{hso}
H_{\rm SO}= -\frac{e}{2 m^\ast}\left[\boldsymbol{\mathbb{A}}(\mathbf{r})\cdot\mathbf{p} + {\rm h.c.}\right]\;.
\end{equation}
Equations (\ref{hamiltonian}) - (\ref{hso}) reveal that a magnetic texture can generate both synthetic SOC and collinear Zeeman interaction, which combined with superconductivity can lead to the emergence of MBS. This observation has motivated 
various proposals for generating MBS in 1D-like systems \cite{Braunecker2010:PRB,Klinovaja2013:PRL,Vazifeh2013:PRL,Marra2017:PRB}. Furthermore, it has recently been shown that the magnetic texture can also provide confinement, opening the possibility of realizing and manipulating MBS in 2D systems \cite{Fatin2016:PRL}.
This is particularly attractive because it can lead to highly localized states without the need for physical wires or dots. A simple example of magnetically-induced confinement is to consider  a texture of the form $\mathbf{J}(\mathbf{r})=J(\mathbf{r})\hat{\mathbf{z}}$. In such a case spin is a good quantum number and the system Hamiltonian reduces to
\begin{equation}\label{hamiltonian-sz}
H = \left(\frac{p^2}{2m^\ast} -\mu \right) + \sigma\;J(\mathbf{r})\sigma_z\;,
\end{equation}
where $\sigma = \pm 1$ for up and down spins, respectively. If the texture is such that $J(\mathbf{r})$ has a maximum (minimum) at a point $\mathbf{r}_0$ and $J(\mathbf{r}\rightarrow \infty)\rightarrow 0$, the Zeeman term forms a confinement potential for spin down (up) particles [see Fig.~\ref{fig:texture}(b)].

In the presence of proximity-induced superconductivity the 2DEG under the effect of the magnetic texture is described by the Bogoliubov-de-Gennes (BdG) Hamiltonian, which in the rotated frame reads as,
\begin{equation}\label{hamiltonian-bdg}
H'_{\rm BdG} = \left(\frac{\left[\mathbf{p}-e\boldsymbol{\mathbb{A}}(\mathbf{r})\right]^2}{2m^\ast}-\mu\right)\tau_z + \Delta \tau_x + J(\mathbf{r}) \sigma_z\;,
\end{equation}
where $\Delta$ is the proximity-induced superconducting gap, and $\tau_i$ are the Nambu matrices in particle-hole space.

\begin{figure}
	\centering
	\includegraphics[width=8.5cm]{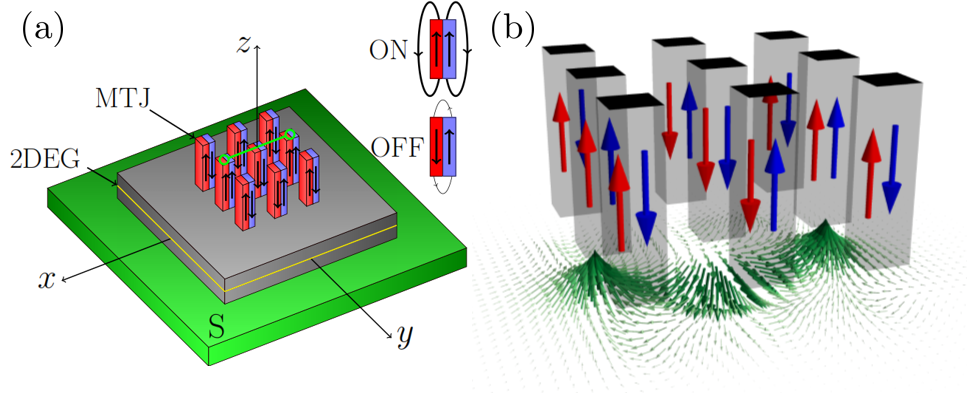}
	\caption{(a) Schematic of the setup. A two-dimensional electron gas (2DEG) is formed in a semiconductor quantum well grown on the surface of an $s$-wave superconductor (S). An array of magnetic tunnel junctions (MTJs) produces a magnetic texture, tunable by switching individual MTJs to the parallel (ON) or antiparallel (OFF) configuration. For the depicted array configuration, two Majorana bound states form at the ends of the middle row (the green line). (b) Magnetic texture produced by the fringing fields generated by the MTJ array.}\label{fig:system}
\end{figure}

There exist multiple ways of generating magnetic textures at the nanoscale. In particular, helical-like textures created by arrays of nanomagnets \cite{Klinovaja2012:PRL,Kjaergaard2012:PRB}, chains of magnetic atoms \cite{NadjPerge2014:S,Pawlak2016:NPJQI,Pientka2013:PRB,Nakosai2013:PRB,Kim2014:PRB,Schecter2016:PRB}, magnetic skyrmions \cite{Yang2016:PRB}, or a ferromagnetic tip \cite{Xu2015:PRL,Sun2016:PRL} have been shown to support MBS. 
However, the realization of braiding operations demands high control and local tunability of the magnetic textures. This poses challenges to those previous proposals, where the magnetic textures are tunable globally but not locally. 
To address such challenges, we propose to use the magnetic textures produced by the fringing fields generated by an array of MTJs,
shown in Fig.~\ref{fig:system}. The texture acts on a 2DEG with superconductivity induced by proximity to an $s$-wave superconductor. The highly inhomogeneous magnetic texture can be tuned by switching individual MTJs to the parallel or antiparallel configuration. The fringing field of a MTJ in the antiparallel configuration is negligible compared to that produced in the parallel state (see Fig.~\ref{fig:system}). Therefore switching to the parallel (antiparallel) configuration turns ON (OFF) the magnetic texture underneath the addressed MTJ. This provides a highly controllable way for locally modifying the magnetic textures acting on the superconducting 2DEG. 

The magnetic texture produced by ON MTJs with alternating magnetization directions leads to topological nontrivial regions enclosed within 
white contours defined by the condition \cite{Fatin2016:PRL},
\begin{equation}\label{t-condition}
|\mathbf{J}(\mathbf{r})|^2=\left[\mu - \eta(\mathbf{r})\right]^2+ \Delta^2\;,
\end{equation}
where
\begin{equation}\label{eta-def}
\eta(\mathbf{r}) = \frac{\hbar^2}{8 m^\ast|\mathbf{J}(\mathbf{r})|^2} \sum_{i=1}^2\left(\frac{\partial \mathbf{J}(\mathbf{r})}{\partial x_i}\cdot \frac{\partial \mathbf{J}(\mathbf{r})}{\partial x_i}\right)\;,
\end{equation}
represents an effective shift in the chemical potential $\mu$ due to local changes of the magnetic texture. Note than in the limit of a homogeneous magnetic field, $\eta \rightarrow 0$ and Eq.~(\ref{t-condition}) reduces to the well-known condition determining the topological phase transition in quantum wires  and rings
\cite{Oreg2010:PRL,Lutchyn2010:PRL,Prada2012:PRB,Franz2013:NN,Scharf2015:PRB}. For a helical texture $\eta\rightarrow \hbar^2 q^2/8m^{\ast}$ and one recovers from Eq.~(\ref{t-condition}) the condition previously reported in Ref.~\onlinecite{Kjaergaard2012:PRB}.
The topological condition for a disordered Rashba wire with proximity induced s-wave superconductivity can be written in a form similar to Eq.~(\ref{t-condition}) but with $\eta$ depending on the disorder strength \cite{Adagideli2014:PRB}.

\begin{figure}
	\centering
	\includegraphics*[width=8.5cm]{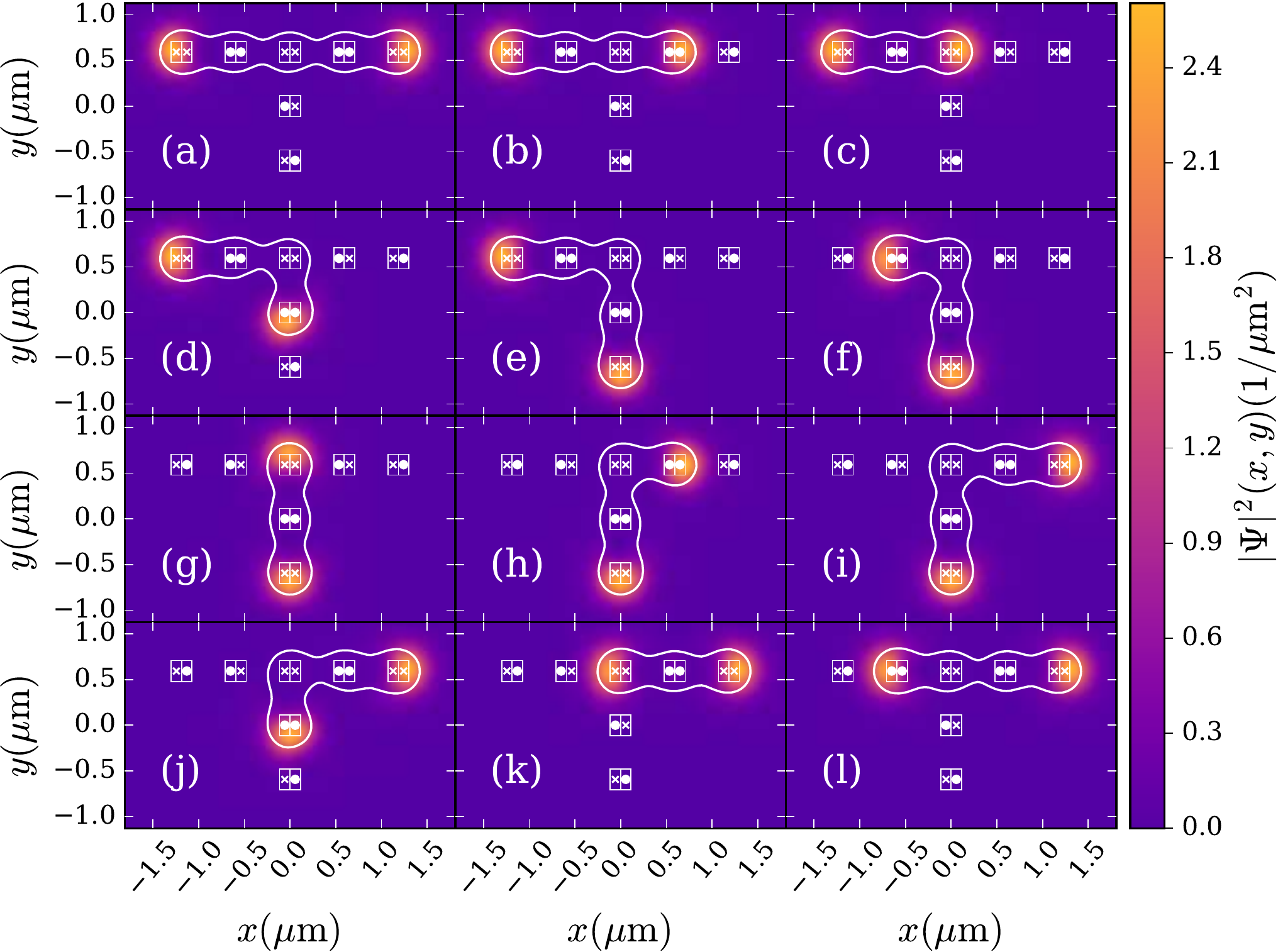}
	\caption{Exchange of two MBS by using a T-shape MTJ array. Color map of the probability densities of the MBS. White rectangles represent the MTJs with dots (crosses) indicating magnetization parallel (antiparallel) to the $\mathbf{z}$ axis. The white contours define reconfigurable effective wires with two end MBS. The sequence of operations from (a) to (l) and back to (a) leads to the physical exchange of the MBS.}\label{fig:t-shape}
\end{figure}

As an illustration we show in Fig.~\ref{fig:t-shape} the exchange of two MBS for the case of an InAs$_{0.4}$Sb$_{0.6}$ 2DEG \cite{Svensson2012:PRB} 
 with proximity induced superconductivity under a T-shape MTJ array. The topological contours define effective wires with MBS localized at the ends. Unlike physical wires, the effective wires are reconfigurable, i.e., their size and shape can be changed by switching individual MTJs. This provides a tunable tool for controlling the position of the MBS as well as their transport and exchange, described in Fig.~\ref{fig:t-shape} for a T-shape MTJ array.

From a practical point of view, 
the potential advantages of the proposed MTJ-based platform can be summarized as follows: (i) no restrictive geometries such as physical wires are required, (ii) the magnetic textures can be locally controlled by electrical switching of individual MTJs (i.e., no external magnetic fields are needed) and (iii) no contacts are required for manipulating MBS, minimizing the risk of quasiparticle poisoning. 
Spin-valves, such as MTJs,  are also the key elements in many spintronic devices, including  magnetic computer hard 
drives~\cite{Tsymbal:2011,Zutic2004:RMP} offering thus the transfer of technological advances to the quest for MBS. In fact the versatility of tunable 
magnetic textures using spin-valves has already been demonstrated in the normal state properties by realizing a spin transistor in  
(Cd,Mn)Te 2DEG~\cite{Betthausen2012:S,Zutic2012:S}.

\vspace{-.2cm}
\subsection{3. Braiding Majorana Bound States}
\vspace{-.2cm}
\begin{figure*}
	\centering
	\includegraphics*[width=\textwidth]{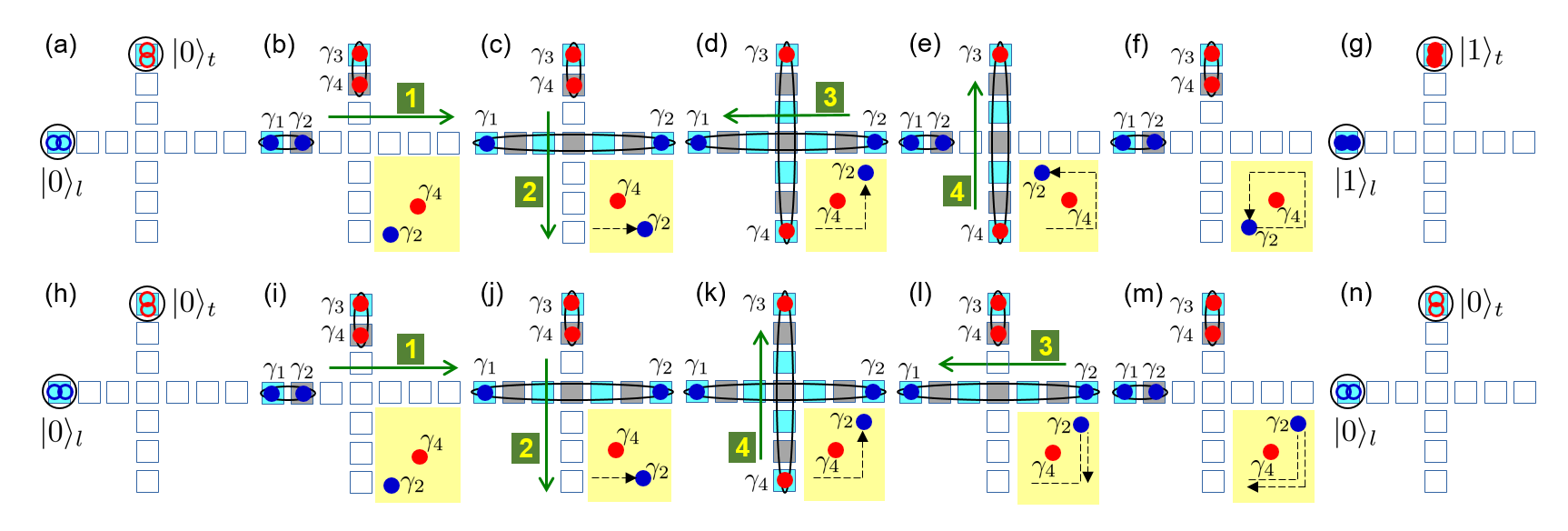}
	\caption{Braiding scheme for probing non-Abelian statistics of MBS. Ligth blue (gray) squares represent MTJs in the ON states with magnetization parallel (antiparallel) to the $\mathbf{z}$ axis, while white squares correspond to MTJs in the OFF state. Black contours
show  the reconfigurable confining structure and full (empty) dots represent occupied (unoccupied) MBS. The sequence (a) - (g) illustrates the adiabatic evolution from two initially empty (neutral) to occupied (charged) QDs upon performing a double braiding. The double braiding is topologically equivalent to the $\gamma_2$ MBS performing a closed path around $\gamma_4$. This is schematically shown in the bottom-right insets of (b) - (f), where the trajectory of $\gamma_2$, as seen in the reference frame of $\gamma_4$, is depicted. The sequence (h) - (n) shows a control experiment similar to (a) - (g) but with the order of steps 3 and 4 inverted. In such a case there is no braiding but a trivial double exchange of MBS [see insets in (i) - (m)] and the QDs remain uncharged.}
	\label{fig:braiding}
\end{figure*}

Beyond the creation, manipulation, and detection of MBS, the ultimate goal is to probe their non-Abelian statistics. 
This could be experimentally realized by using the MTJ-array platform for performing braiding operations and measuring the changes in local charge density. Probing the non-Abelian statistics requires the manipulation of at least 4 MBS. As a proof of principle,
we use a 9-MTJs array in Fig.~\ref{fig:braiding}. The MTJ-array is initially prepared in a configuration with only the left and top most MTJs in the ON state 
[see Fig.~\ref{fig:braiding}(a)]. The corresponding magnetic texture leads to the formation of two effective quantum dots (QDs). The QDs do not support the MBS formation and therefore 
their confined states correspond to finite-energy excitations. The lowest energy levels of the left and top QDs are initially unoccupied [see Fig.~\ref{fig:braiding}(a)].

Since the effective QDs are initially empty the corresponding charge and probability densities vanish. By switching the two adjacent MTJs, the QDs adiabatically evolves into topological quantum wires with two occupied \emph{zero-energy} MBS each [$\gamma_1$, $\gamma_2$ and $\gamma_3$, $\gamma_4$ in Fig.~\ref{fig:braiding}(b)]. The charge density in the wires remains close to zero because the MBS are nearly chargeless.

The subsequent switching of the MTJs along the directions indicated by the green arrows in Figs.~\ref{fig:braiding}(b) and (c), steps 1 and 2 respectively, yields the formation of two crossing effective wires, as shown in Fig.~\ref{fig:braiding}(d). Performing steps 1 and 2 is equivalent to an anticlockwise exchange of the MBS $\gamma_2$ and $\gamma_4$. Furthermore, an additional braiding of the MBS $\gamma_2$ and $\gamma_4$ is realized by performing steps 3 and 4, as indicated in Figs.~\ref{fig:braiding}(d) and (e). The double braiding is topologically equivalent to one MBS, say $\gamma_2$, performing a closed path around $\gamma_4$, as illustrated in the bottom-right insets in Figs.~\ref{fig:braiding}(b) - (f).

Turning back the effective wires into QDs [see Figs.~\ref{fig:braiding}(f) and (g)] causes the MBS to fuse. Therefore, the complete operation in which the system evolves from Fig.~\ref{fig:braiding}(a) to (g) corresponds to a double braiding in which the initial state of the empty QDs adiabatically evolves into a final state with the two QDs being occupied. This leads to the emergence of extra charge in the effective QDs after the double braiding. 

The non-Abelian character of the MBS manifests when comparing the order of operations. For example, a complementary control experiment in which the order of operations 3 and 4 is inverted [see Figs.~\ref{fig:braiding}(h) - (n)] leads to a trivial double exchange and no extra charge emerges in the dots. Thus, by contrasting the initial and final charges of the QDs under the different sequences of operations the non-Abelian character of the statistics obeyed by the MBS can be measured. Numerical simulations illustrating the proposed braiding scheme for probing the non-Abelian statistics of MBS  were based on the parameters of 
(Cd,Mn)Te 2DEG~\cite{Betthausen2012:S} 
with proximity induced superconductivity. They have shown the transition from the initially empty to charged QDs upon 
performing a double braiding \cite{Fatin2016:PRL}. 
However, as shown in Fig.~\ref{fig:t-shape}, 
MBS can also be created and manipulated in other high $g$-factor materials such as 
InAs$_{0.4}$Sb$_{0.6}$ \cite{Svensson2012:PRB}.

The outlined braiding protocol in Fig.~\ref{fig:braiding} is also important in distinguishing possible competing effects that could be attributed to MBS. A  typical  example is the formation of topologically-trivial Andreev bound states (ABS) that can lead to zero bias conductance peak (ZBCP) due to spatial inhomogeneity of the superconducting pair potential for various pairing symmetries, as well as in junctions with magnetic 
materials~\cite{Sengupta2001:PRB,Hu1994:PRL,Kashiwaya1995:PRL,Wei1998:PRL,Kashiwaya2000:RPP,Zutic2000:PRB,Chen2001:PRB}. 
While the completion of double braiding of MBS depicted in Fig.~\ref{fig:braiding} leads to the difference between the initial and final charge of the QD, it does not affect ABS. Therefore, the non-Abelian character reflected in the different outcomes of the braiding protocol, for which ABS play no role, is a direct and unambiguous signature of MBS.


A less direct way of trying to disentangle ABS vs MBS contributions would be to look at potential differences in the evolution of the ZBCP with temperature, magnetic field, and even interface orientation. Most of the efforts have so far focused on considering a native $s$-wave superconductor where such a distinction remains difficult~\cite{Deng2016:S,Liu2017:P,Liu2012:PRL}, while some additional possibilities could be possible for $d$-wave symmetry where the interfacial orientation between the normal region and a superconductor plays an important role~\cite{Tsuei2013:Arxiv}.

\vspace{-.2cm}
\subsection{4. Theoretical Opportunities and Experimental Challenges}
\vspace{-.2cm}
To assess the prospect of employing tunable magnetic textures for manipulating MBS, it is helpful to summarize some of the advantages of this platform, as well as to identify the key challenges. Unlike the most frequent MBS implementations based on semiconductor nanowires, neither the applied field, nor strong SOC are required. No electrostatic contacts are needed for creating and manipulating MBS therefore minimizing the effects of quasiparticle poisoning. While magnetic textures are also employed in magnetic atomic chains, in the present proposal the magnetic textures are highly-tunable, allowing for a versatile control of MBS. Our 2D platform overcomes various limitations for braiding, inherent to the 1D geometry of physical wires, and offers a path to scalable MBS.

\begin{figure}
	\centering
	\includegraphics*[width=8.5cm]{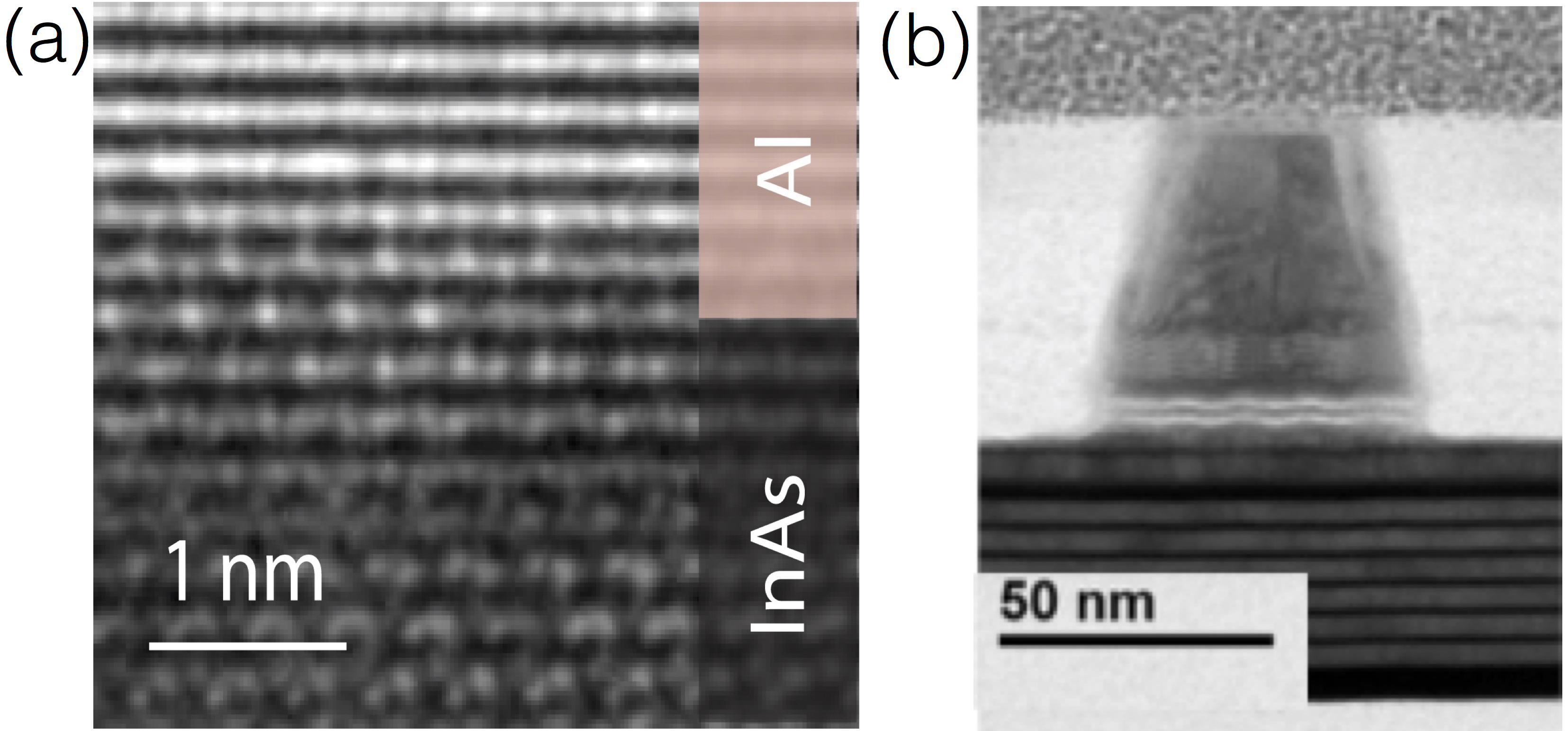}
	\caption{(a) Epitaxial growth of superconducting Al film on semiconducting InAs quantum well. (b) Scanning transmission electron
		microscope image of a MTJ nanopillar. The MgO tunnel barrier is visible as the lower bright line. This separates the free and fixed magnetic layer.}
	\label{Fig:InAsAl_MTJ}
\end{figure}

Although individual elements of our platform, MTJs, superconductors, and semiconductors, are extensively studied, combining them in a 2D geometry that maintains robust superconducting proximity effects remains a challenging task. We have recently demonstrated an important step towards integrating these elements by epitaxially growing ultra-thin Al films on InAs surface quantum wells \cite{Shabani2016:PRB} and
signatures of MBS on these heterostructures have been found \cite{Suominen2017:P}. A transmission electron microscope image of an Al/InAs epitaxial interface is shown in Fig.~\ref{Fig:InAsAl_MTJ}. 

Al thin films have a relatively small superconducting gap ($\sim 200$ $\mu$eV) and critical magnetic fields ($B_{c} \sim 100$ mT in perpendicular and up to 2 T in parallel direction). On the other hand, Al has a very long coherence length that is desired in most superconducting devices. In our proposal materials with large $g$-factor will lower the magnetic field required for the transition into the topological regime and hence are desired. One such family is InAs$_{1-x}$Sb$_{x}$.  The $g$-factor in InAs is about $g \sim$ -10 (at $x = $ 0) while this magnitude is increased to $g \sim$ -50 in InSb ($x =$ 1)~\cite{Zutic2004:RMP,Winkler2017:P}. Due to the bowing of the bandgap near $x \sim 0.6$, the $g$-factor is enhanced even more than InSb to $g \sim$ -140 \cite{Svensson2012:PRB}. Strain energy calculations suggest that similar interfaces can be achieved in Al/InAs$_{0.4}$Sb$_{0.6}$ material systems.

To better understand the implementation of the topological condition for effective wires in Eq.~(\ref{t-condition}), we focus on
the InAs-based system already used as a promising 2D platform for MBS~\cite{Shabani2016:PRB}.  
The density of InAs structures are determined by the position of the Fermi level pinning at the surface. Indeed, for a pure InAs this results 
in a rather high density of carriers making it harder to control and tune the density to yield a sufficiently small chemical potential, $\mu$,
needed to satisfy the topological condition required for MBS. However, 
we have  recently shown 
in Ref.~\cite{Shabani2016:PRB} that the Fermi level pinning position can be modified by introduction of In$_{x}$Ga$_{1-x}$As capping layer. 
At x = 0, GaAs, the Fermi level is pinned in the band gap resulting in no free carrier near the surface. The Fermi level pinning will cross the 
conduction band near  $x = 0.85$ meaning that we expect zero density at $T = 0$  K for $x < 0.85$ and monotonically increasing by increasing 
$x$ above 0.85. Considering a small lattice mismatch in this range of $x$, the growth of few nanometers of (In,Ga)As films is possible and has been
demonstrated~\cite{Shabani2016:PRB}. This tuning can allow controlling the magnitude of $\mu$. 
Based on the realized carrier densities of $\lesssim 10^{11}$ cm$^{-2}$, the corresponding chemical potential is $\mu \sim 1$ meV. 
This approach supports the experimental feasibility of the topological condition needed to observe MBS.
Note that the effective chemical potential, $\mu - \eta$, entering in the topological condition is further modified by the shift $\eta$, which can be controlled through magnetic textures realized by built-in magnets.
If needed, an additional control can be achieved by adding a constant magnetic field which changes the left-hand-side of Eq.~(\ref{t-condition}), while leaving the 
right-hand-side unchanged ($\eta$ is given by the gradient of the texture and remains the same after adding a constant magnetic field). Thus, by gradually increasing the external field, one can tune the system such that the topological condition in Eq.~(\ref{t-condition}) is reached.

Another experimental challenge is to identify a suitable implementation of tunable magnetic textures that could be integrated
with proximity-induced superconductivity in the 2DEG. Such magnetic textures can be created using arrays of MTJ nanopillars that consist of fixed and free magnetic layers. The layer magnetizations can be oriented either parallel or antiparrallel to turn on or greatly reduce the fringing field. Further, the magnetic textures can be tuned and changed dynamically using spin-transfer torques (STT), as implemented in commercial magnetic random access memories (STT-MRAM)~\cite{Kent2015:NN}.  A great deal of progress has been made in the past decade driven by the development of magnetic random access memories for the semiconductor industry. Although switching magnetic domains with magnetic fields is relatively slow, it has been shown that spin-transfer torque (STT) devices allow a fully electrical control over the magnetic texture, and nanopillars states can be changed on sub-ns time scales \cite{Hahn2016:PRB}. STT controlled MTJ nanopillars as small as 11 nm in diameter have been realized \cite{Nowak2016:IEEEML} 
and large scale arrays of 20 nm diameter nanopillars are feasible with a pitch 4 times their diameter. As the main requirement for MBS
formation and braiding is setting and controllably modifying the magnetic texture, the nanopillars can also be formed from all metallic spin-valve structures that can switch at similar current densities but with much lower dissipation \cite{Ye2014:JAP}, because of their lower impedance, 
$\sim$ $\Omega$ vs k$\Omega$
for STT-switchable MTJs at the 20 nm diameter size scale.

\vspace{-.2cm}
\subsection{5. Conclusions}
\vspace{-.2cm}
In the current situation where there is a strong support for the experimental realization of MBS, but the key proof of non-Abelian statistics through braiding is still missing, the present 2D platform using tunable magnetic textures provides a viable alternative to the widely pursued 1D geometries. Moreover, supported by the recent experimental advances in realizing such a 2D geometry, it could be possible to use our platform to also implement other topological states, beyond MBS.\\

\noindent\emph{Acknowledgments.} This work was supported by U.S. DOE, Office of Science BES, under Award ${\rm DE}$-SC0004890
(A.M.-A., G.L.F., B.S., I.\v{Z}), DFG Grants no. SCHA 1899/1-1 and SFB 1170 "ToCoTronics" and the ENB Graduate School on Topological Insulators (B.S.), NSF DGE-1650441 (G.L.F.). 
J.S. acknowledges support from Army Research Office and Air Force Office of Scientific Research.
The research at NYU was supported in part by the National Science Foundation under Award DMR-1610416.


\bibliographystyle{apsrev4-1}

\begin{thebibliography}{73}%
	\makeatletter
	\providecommand \@ifxundefined [1]{%
		\@ifx{#1\undefined}
	}%
	\providecommand \@ifnum [1]{%
		\ifnum #1\expandafter \@firstoftwo
		\else \expandafter \@secondoftwo
		\fi
	}%
	\providecommand \@ifx [1]{%
		\ifx #1\expandafter \@firstoftwo
		\else \expandafter \@secondoftwo
		\fi
	}%
	\providecommand \natexlab [1]{#1}%
	\providecommand \enquote  [1]{``#1''}%
	\providecommand \bibnamefont  [1]{#1}%
	\providecommand \bibfnamefont [1]{#1}%
	\providecommand \citenamefont [1]{#1}%
	\providecommand \href@noop [0]{\@secondoftwo}%
	\providecommand \href [0]{\begingroup \@sanitize@url \@href}%
	\providecommand \@href[1]{\@@startlink{#1}\@@href}%
	\providecommand \@@href[1]{\endgroup#1\@@endlink}%
	\providecommand \@sanitize@url [0]{\catcode `\\12\catcode `\$12\catcode
		`\&12\catcode `\#12\catcode `\^12\catcode `\_12\catcode `\%12\relax}%
	\providecommand \@@startlink[1]{}%
	\providecommand \@@endlink[0]{}%
	\providecommand \url  [0]{\begingroup\@sanitize@url \@url }%
	\providecommand \@url [1]{\endgroup\@href {#1}{\urlprefix }}%
	\providecommand \urlprefix  [0]{URL }%
	\providecommand \Eprint [0]{\href }%
	\providecommand \doibase [0]{http://dx.doi.org/}%
	\providecommand \selectlanguage [0]{\@gobble}%
	\providecommand \bibinfo  [0]{\@secondoftwo}%
	\providecommand \bibfield  [0]{\@secondoftwo}%
	\providecommand \translation [1]{[#1]}%
	\providecommand \BibitemOpen [0]{}%
	\providecommand \bibitemStop [0]{}%
	\providecommand \bibitemNoStop [0]{.\EOS\space}%
	\providecommand \EOS [0]{\spacefactor3000\relax}%
	\providecommand \BibitemShut  [1]{\csname bibitem#1\endcsname}%
	\let\auto@bib@innerbib\@empty
	\bibitem [{\citenamefont {Nayak}\ \emph {et~al.}(2008)\citenamefont {Nayak},
		\citenamefont {Simon}, \citenamefont {Stern}, \citenamefont {Freedman},\ and\
		\citenamefont {Das~Sarma}}]{Nayak2008:RMP}%
	\BibitemOpen
	\bibfield  {author} {\bibinfo {author} {\bibfnamefont {C.}~\bibnamefont
			{Nayak}}, \bibinfo {author} {\bibfnamefont {S.~H.}\ \bibnamefont {Simon}},
		\bibinfo {author} {\bibfnamefont {A.}~\bibnamefont {Stern}}, \bibinfo
		{author} {\bibfnamefont {M.}~\bibnamefont {Freedman}}, \ and\ \bibinfo
		{author} {\bibfnamefont {S.}~\bibnamefont {Das~Sarma}},\ }\href {\doibase
		10.1103/RevModPhys.80.1083} {\bibfield  {journal} {\bibinfo  {journal} {Rev.
				Mod. Phys.}\ }\textbf {\bibinfo {volume} {80}},\ \bibinfo {pages} {1083}
		(\bibinfo {year} {2008})}\BibitemShut {NoStop}%
	\bibitem [{\citenamefont {Kitaev}(2003)}]{Kitaev2003:AP}%
	\BibitemOpen
	\bibfield  {author} {\bibinfo {author} {\bibfnamefont {A.}~\bibnamefont
			{Kitaev}},\ }\href {http://stacks.iop.org/1468-6996/15/i=6/a=064402}
	{\bibfield  {journal} {\bibinfo  {journal} {Ann. Phys.}\ }\textbf {\bibinfo
			{volume} {303}},\ \bibinfo {pages} {2} (\bibinfo {year} {2003})}\BibitemShut
	{NoStop}%
	\bibitem [{\citenamefont {Ivanov}(2001)}]{Ivanov2001:PRL}%
	\BibitemOpen
	\bibfield  {author} {\bibinfo {author} {\bibfnamefont {D.~A.}\ \bibnamefont
			{Ivanov}},\ }\href {\doibase 10.1103/PhysRevLett.86.268} {\bibfield
		{journal} {\bibinfo  {journal} {Phys. Rev. Lett.}\ }\textbf {\bibinfo
			{volume} {86}},\ \bibinfo {pages} {268} (\bibinfo {year} {2001})}\BibitemShut
	{NoStop}%
	\bibitem [{\citenamefont {Mackenzie}\ and\ \citenamefont
		{Maeno}(2003)}]{Mackenzie2003:RMP}%
	\BibitemOpen
	\bibfield  {author} {\bibinfo {author} {\bibfnamefont {A.~P.}\ \bibnamefont
			{Mackenzie}}\ and\ \bibinfo {author} {\bibfnamefont {Y.}~\bibnamefont
			{Maeno}},\ }\href {\doibase 10.1103/RevModPhys.75.657} {\bibfield  {journal}
		{\bibinfo  {journal} {Rev. Mod. Phys.}\ }\textbf {\bibinfo {volume} {75}},\
		\bibinfo {pages} {657} (\bibinfo {year} {2003})}\BibitemShut {NoStop}%
	\bibitem [{\citenamefont {Sengupta}\ \emph {et~al.}(2001)\citenamefont
		{Sengupta}, \citenamefont {\ifmmode \check{Z}\else
			\v{Z}\fi{}uti\ifmmode~\acute{c}\else \'{c}\fi{}}, \citenamefont {Kwon},
		\citenamefont {Yakovenko},\ and\ \citenamefont
		{Das~Sarma}}]{Sengupta2001:PRB}%
	\BibitemOpen
	\bibfield  {author} {\bibinfo {author} {\bibfnamefont {K.}~\bibnamefont
			{Sengupta}}, \bibinfo {author} {\bibfnamefont {I.}~\bibnamefont {\ifmmode
				\check{Z}\else \v{Z}\fi{}uti\ifmmode~\acute{c}\else \'{c}\fi{}}}, \bibinfo
		{author} {\bibfnamefont {H.-J.}\ \bibnamefont {Kwon}}, \bibinfo {author}
		{\bibfnamefont {V.~M.}\ \bibnamefont {Yakovenko}}, \ and\ \bibinfo {author}
		{\bibfnamefont {S.}~\bibnamefont {Das~Sarma}},\ }\href {\doibase
		10.1103/PhysRevB.63.144531} {\bibfield  {journal} {\bibinfo  {journal} {Phys.
				Rev. B}\ }\textbf {\bibinfo {volume} {63}},\ \bibinfo {pages} {144531}
		(\bibinfo {year} {2001})}\BibitemShut {NoStop}%
	\bibitem [{\citenamefont {Wu}\ and\ \citenamefont
		{Samokhin}(2010)}]{Wu2010:PRB}%
	\BibitemOpen
	\bibfield  {author} {\bibinfo {author} {\bibfnamefont {S.}~\bibnamefont
			{Wu}}\ and\ \bibinfo {author} {\bibfnamefont {K.~V.}\ \bibnamefont
			{Samokhin}},\ }\href {\doibase 10.1103/PhysRevB.81.214506} {\bibfield
		{journal} {\bibinfo  {journal} {Phys. Rev. B}\ }\textbf {\bibinfo {volume}
			{81}},\ \bibinfo {pages} {214506} (\bibinfo {year} {2010})}\BibitemShut
	{NoStop}%
	\bibitem [{\citenamefont {\ifmmode \check{Z}\else
			\v{Z}\fi{}uti\ifmmode~\acute{c}\else \'{c}\fi{}}\ and\ \citenamefont
		{Mazin}(2005)}]{Zutic2005:PRL}%
	\BibitemOpen
	\bibfield  {author} {\bibinfo {author} {\bibfnamefont {I.}~\bibnamefont
			{\ifmmode \check{Z}\else \v{Z}\fi{}uti\ifmmode~\acute{c}\else \'{c}\fi{}}}\
		and\ \bibinfo {author} {\bibfnamefont {I.}~\bibnamefont {Mazin}},\ }\href
	{\doibase 10.1103/PhysRevLett.95.217004} {\bibfield  {journal} {\bibinfo
			{journal} {Phys. Rev. Lett.}\ }\textbf {\bibinfo {volume} {95}},\ \bibinfo
		{pages} {217004} (\bibinfo {year} {2005})}\BibitemShut {NoStop}%
	\bibitem [{\citenamefont {Fu}\ and\ \citenamefont {Kane}(2008)}]{Fu2008:PRL}%
	\BibitemOpen
	\bibfield  {author} {\bibinfo {author} {\bibfnamefont {L.}~\bibnamefont
			{Fu}}\ and\ \bibinfo {author} {\bibfnamefont {C.~L.}\ \bibnamefont {Kane}},\
	}\href {\doibase 10.1103/PhysRevLett.100.096407} {\bibfield  {journal}
	{\bibinfo  {journal} {Phys. Rev. Lett.}\ }\textbf {\bibinfo {volume} {100}},\
	\bibinfo {pages} {096407} (\bibinfo {year} {2008})}\BibitemShut {NoStop}%
\bibitem [{\citenamefont {Lutchyn}\ \emph {et~al.}(2010)\citenamefont
	{Lutchyn}, \citenamefont {Sau},\ and\ \citenamefont
	{Das~Sarma}}]{Lutchyn2010:PRL}%
\BibitemOpen
\bibfield  {author} {\bibinfo {author} {\bibfnamefont {R.~M.}\ \bibnamefont
		{Lutchyn}}, \bibinfo {author} {\bibfnamefont {J.~D.}\ \bibnamefont {Sau}}, \
	and\ \bibinfo {author} {\bibfnamefont {S.}~\bibnamefont {Das~Sarma}},\ }\href
{\doibase 10.1103/PhysRevLett.105.077001} {\bibfield  {journal} {\bibinfo
		{journal} {Phys. Rev. Lett.}\ }\textbf {\bibinfo {volume} {105}},\ \bibinfo
	{pages} {077001} (\bibinfo {year} {2010})}\BibitemShut {NoStop}%
\bibitem [{\citenamefont {Sau}\ \emph {et~al.}(2010)\citenamefont {Sau},
	\citenamefont {Lutchyn}, \citenamefont {Tewari},\ and\ \citenamefont
	{Das~Sarma}}]{Sau2010:PRL}%
\BibitemOpen
\bibfield  {author} {\bibinfo {author} {\bibfnamefont {J.~D.}\ \bibnamefont
		{Sau}}, \bibinfo {author} {\bibfnamefont {R.~M.}\ \bibnamefont {Lutchyn}},
	\bibinfo {author} {\bibfnamefont {S.}~\bibnamefont {Tewari}}, \ and\ \bibinfo
	{author} {\bibfnamefont {S.}~\bibnamefont {Das~Sarma}},\ }\href {\doibase
	10.1103/PhysRevLett.104.040502} {\bibfield  {journal} {\bibinfo  {journal}
		{Phys. Rev. Lett.}\ }\textbf {\bibinfo {volume} {104}},\ \bibinfo {pages}
	{040502} (\bibinfo {year} {2010})}\BibitemShut {NoStop}%
\bibitem [{\citenamefont {Oreg}\ \emph {et~al.}(2010)\citenamefont {Oreg},
	\citenamefont {Refael},\ and\ \citenamefont {von Oppen}}]{Oreg2010:PRL}%
\BibitemOpen
\bibfield  {author} {\bibinfo {author} {\bibfnamefont {Y.}~\bibnamefont
		{Oreg}}, \bibinfo {author} {\bibfnamefont {G.}~\bibnamefont {Refael}}, \ and\
	\bibinfo {author} {\bibfnamefont {F.}~\bibnamefont {von Oppen}},\ }\href
{\doibase 10.1103/PhysRevLett.105.177002} {\bibfield  {journal} {\bibinfo
		{journal} {Phys. Rev. Lett.}\ }\textbf {\bibinfo {volume} {105}},\ \bibinfo
	{pages} {177002} (\bibinfo {year} {2010})}\BibitemShut {NoStop}%
\bibitem [{\citenamefont {Alicea}(2012)}]{Alicea2012:RPP}%
\BibitemOpen
\bibfield  {author} {\bibinfo {author} {\bibfnamefont {J.}~\bibnamefont
		{Alicea}},\ }\href@noop {} {\bibfield  {journal} {\bibinfo  {journal} {Rep.
			Prog. Phys.}\ }\textbf {\bibinfo {volume} {75}},\ \bibinfo {pages} {076501}
	(\bibinfo {year} {2012})}\BibitemShut {NoStop}%
\bibitem [{\citenamefont {Leijnse}\ and\ \citenamefont
	{Flensberg}(2012)}]{Leijnse2012:SST}%
\BibitemOpen
\bibfield  {author} {\bibinfo {author} {\bibfnamefont {M.}~\bibnamefont
		{Leijnse}}\ and\ \bibinfo {author} {\bibfnamefont {K.}~\bibnamefont
		{Flensberg}},\ }\href@noop {} {\bibfield  {journal} {\bibinfo  {journal}
		{Semicond. Sci. and Technol.}\ }\textbf {\bibinfo {volume} {27}},\ \bibinfo
	{pages} {124003} (\bibinfo {year} {2012})}\BibitemShut {NoStop}%
\bibitem [{\citenamefont {Mourik}\ \emph {et~al.}(2012)\citenamefont {Mourik},
	\citenamefont {Zuo}, \citenamefont {Frolov}, \citenamefont {Plissard},
	\citenamefont {Bakkers},\ and\ \citenamefont {Kouwenhoven}}]{Mourik2012:S}%
\BibitemOpen
\bibfield  {author} {\bibinfo {author} {\bibfnamefont {V.}~\bibnamefont
		{Mourik}}, \bibinfo {author} {\bibfnamefont {K.}~\bibnamefont {Zuo}},
	\bibinfo {author} {\bibfnamefont {S.~M.}\ \bibnamefont {Frolov}}, \bibinfo
	{author} {\bibfnamefont {S.~R.}\ \bibnamefont {Plissard}}, \bibinfo {author}
	{\bibfnamefont {E.~P. A.~M.}\ \bibnamefont {Bakkers}}, \ and\ \bibinfo
	{author} {\bibfnamefont {L.~P.}\ \bibnamefont {Kouwenhoven}},\ }\href@noop {}
{\bibfield  {journal} {\bibinfo  {journal} {Science}\ }\textbf {\bibinfo
		{volume} {336}},\ \bibinfo {pages} {1003} (\bibinfo {year}
	{2012})}\BibitemShut {NoStop}%
\bibitem [{\citenamefont {Deng}\ \emph {et~al.}(2012)\citenamefont {Deng},
	\citenamefont {Yu}, \citenamefont {Huang}, \citenamefont {Larsson},
	\citenamefont {Caroff},\ and\ \citenamefont {Xu}}]{Deng2012:NL}%
\BibitemOpen
\bibfield  {author} {\bibinfo {author} {\bibfnamefont {M.~T.}\ \bibnamefont
		{Deng}}, \bibinfo {author} {\bibfnamefont {C.~L.}\ \bibnamefont {Yu}},
	\bibinfo {author} {\bibfnamefont {G.~Y.}\ \bibnamefont {Huang}}, \bibinfo
	{author} {\bibfnamefont {M.}~\bibnamefont {Larsson}}, \bibinfo {author}
	{\bibfnamefont {P.}~\bibnamefont {Caroff}}, \ and\ \bibinfo {author}
	{\bibfnamefont {H.~Q.}\ \bibnamefont {Xu}},\ }\href {\doibase
	10.1021/nl303758w} {\bibfield  {journal} {\bibinfo  {journal} {Nano Letters}\
	}\textbf {\bibinfo {volume} {12}},\ \bibinfo {pages} {6414} (\bibinfo {year}
	{2012})}\BibitemShut {NoStop}%
\bibitem [{\citenamefont {Rokhinson}\ \emph {et~al.}(2012)\citenamefont
	{Rokhinson}, \citenamefont {Liu},\ and\ \citenamefont
	{Furdyna}}]{Rokhinson2012:NP}%
\BibitemOpen
\bibfield  {author} {\bibinfo {author} {\bibfnamefont {L.~P.}\ \bibnamefont
		{Rokhinson}}, \bibinfo {author} {\bibfnamefont {X.}~\bibnamefont {Liu}}, \
	and\ \bibinfo {author} {\bibfnamefont {J.~K.}\ \bibnamefont {Furdyna}},\
}\href {\doibase 10.1038/nphys2429} {\bibfield  {journal} {\bibinfo
	{journal} {Nature Phys.}\ }\textbf {\bibinfo {volume} {8}},\ \bibinfo {pages}
{795} (\bibinfo {year} {2012})}\BibitemShut {NoStop}%
\bibitem [{\citenamefont {Das}\ \emph {et~al.}(2012)\citenamefont {Das},
	\citenamefont {Ronen}, \citenamefont {Most}, \citenamefont {Oreg},
	\citenamefont {Heiblum},\ and\ \citenamefont {Shtrikman}}]{Das2012:NP}%
\BibitemOpen
\bibfield  {author} {\bibinfo {author} {\bibfnamefont {A.}~\bibnamefont
		{Das}}, \bibinfo {author} {\bibfnamefont {Y.}~\bibnamefont {Ronen}}, \bibinfo
	{author} {\bibfnamefont {Y.}~\bibnamefont {Most}}, \bibinfo {author}
	{\bibfnamefont {Y.}~\bibnamefont {Oreg}}, \bibinfo {author} {\bibfnamefont
		{M.}~\bibnamefont {Heiblum}}, \ and\ \bibinfo {author} {\bibfnamefont
		{H.}~\bibnamefont {Shtrikman}},\ }\href {\doibase 10.1038/nphys2479}
{\bibfield  {journal} {\bibinfo  {journal} {Nature Phys.}\ }\textbf {\bibinfo
		{volume} {8}},\ \bibinfo {pages} {887} (\bibinfo {year} {2012})}\BibitemShut
{NoStop}%
\bibitem [{\citenamefont {Finck}\ \emph {et~al.}(2013)\citenamefont {Finck},
	\citenamefont {Van~Harlingen}, \citenamefont {Mohseni}, \citenamefont
	{Jung},\ and\ \citenamefont {Li}}]{Finck2013:PRL}%
\BibitemOpen
\bibfield  {author} {\bibinfo {author} {\bibfnamefont {A.~D.~K.}\
		\bibnamefont {Finck}}, \bibinfo {author} {\bibfnamefont {D.~J.}\ \bibnamefont
		{Van~Harlingen}}, \bibinfo {author} {\bibfnamefont {P.~K.}\ \bibnamefont
		{Mohseni}}, \bibinfo {author} {\bibfnamefont {K.}~\bibnamefont {Jung}}, \
	and\ \bibinfo {author} {\bibfnamefont {X.}~\bibnamefont {Li}},\ }\href
{\doibase 10.1103/PhysRevLett.110.126406} {\bibfield  {journal} {\bibinfo
		{journal} {Phys. Rev. Lett.}\ }\textbf {\bibinfo {volume} {110}},\ \bibinfo
	{pages} {126406} (\bibinfo {year} {2013})}\BibitemShut {NoStop}%
\bibitem [{\citenamefont {Nadj-Perge}\ \emph {et~al.}(2014)\citenamefont
	{Nadj-Perge}, \citenamefont {Drozdov}, \citenamefont {Li}, \citenamefont
	{Chen}, \citenamefont {Jeon}, \citenamefont {Seo}, \citenamefont {MacDonald},
	\citenamefont {Bernevig},\ and\ \citenamefont {Yazdani}}]{NadjPerge2014:S}%
\BibitemOpen
\bibfield  {author} {\bibinfo {author} {\bibfnamefont {S.}~\bibnamefont
		{Nadj-Perge}}, \bibinfo {author} {\bibfnamefont {I.~K.}\ \bibnamefont
		{Drozdov}}, \bibinfo {author} {\bibfnamefont {J.}~\bibnamefont {Li}},
	\bibinfo {author} {\bibfnamefont {H.}~\bibnamefont {Chen}}, \bibinfo {author}
	{\bibfnamefont {S.}~\bibnamefont {Jeon}}, \bibinfo {author} {\bibfnamefont
		{J.}~\bibnamefont {Seo}}, \bibinfo {author} {\bibfnamefont {A.~H.}\
		\bibnamefont {MacDonald}}, \bibinfo {author} {\bibfnamefont {B.~A.}\
		\bibnamefont {Bernevig}}, \ and\ \bibinfo {author} {\bibfnamefont
		{A.}~\bibnamefont {Yazdani}},\ }\href {\doibase 10.1126/science.1259327}
{\bibfield  {journal} {\bibinfo  {journal} {Science}\ }\textbf {\bibinfo
		{volume} {346}},\ \bibinfo {pages} {602} (\bibinfo {year}
	{2014})}\BibitemShut {NoStop}%
\bibitem [{\citenamefont {Pawlak}\ \emph {et~al.}(2016)\citenamefont {Pawlak},
	\citenamefont {Kisiel}, \citenamefont {Klinovaja}, \citenamefont {Meier},
	\citenamefont {Kawai}, \citenamefont {Glatzel}, \citenamefont {Loss},\ and\
	\citenamefont {Meyer}}]{Pawlak2016:NPJQI}%
\BibitemOpen
\bibfield  {author} {\bibinfo {author} {\bibfnamefont {R.}~\bibnamefont
		{Pawlak}}, \bibinfo {author} {\bibfnamefont {M.}~\bibnamefont {Kisiel}},
	\bibinfo {author} {\bibfnamefont {J.}~\bibnamefont {Klinovaja}}, \bibinfo
	{author} {\bibfnamefont {T.}~\bibnamefont {Meier}}, \bibinfo {author}
	{\bibfnamefont {S.}~\bibnamefont {Kawai}}, \bibinfo {author} {\bibfnamefont
		{T.}~\bibnamefont {Glatzel}}, \bibinfo {author} {\bibfnamefont
		{D.}~\bibnamefont {Loss}}, \ and\ \bibinfo {author} {\bibfnamefont
		{E.}~\bibnamefont {Meyer}},\ }\href@noop {} {\bibfield  {journal} {\bibinfo
		{journal} {NPJ Quantum Inf.}\ }\textbf {\bibinfo {volume} {2}},\ \bibinfo
	{pages} {16035} (\bibinfo {year} {2016})}\BibitemShut {NoStop}%
\bibitem [{\citenamefont {Deng}\ \emph {et~al.}(2016)\citenamefont {Deng},
	\citenamefont {Vaitiekenas}, \citenamefont {Hansen}, \citenamefont {Danon},
	\citenamefont {Leijnse}, \citenamefont {Flensberg}, \citenamefont {Nyg{\r
			a}rd}, \citenamefont {Krogstrup},\ and\ \citenamefont {Marcus}}]{Deng2016:S}%
\BibitemOpen
\bibfield  {author} {\bibinfo {author} {\bibfnamefont {M.~T.}\ \bibnamefont
		{Deng}}, \bibinfo {author} {\bibfnamefont {S.}~\bibnamefont {Vaitiekenas}},
	\bibinfo {author} {\bibfnamefont {E.~B.}\ \bibnamefont {Hansen}}, \bibinfo
	{author} {\bibfnamefont {J.}~\bibnamefont {Danon}}, \bibinfo {author}
	{\bibfnamefont {M.}~\bibnamefont {Leijnse}}, \bibinfo {author} {\bibfnamefont
		{K.}~\bibnamefont {Flensberg}}, \bibinfo {author} {\bibfnamefont
		{J.}~\bibnamefont {Nyg{\r a}rd}}, \bibinfo {author} {\bibfnamefont
		{P.}~\bibnamefont {Krogstrup}}, \ and\ \bibinfo {author} {\bibfnamefont
		{C.~M.}\ \bibnamefont {Marcus}},\ }\href {\doibase 10.1126/science.aaf3961}
{\bibfield  {journal} {\bibinfo  {journal} {Science}\ }\textbf {\bibinfo
		{volume} {354}},\ \bibinfo {pages} {1557} (\bibinfo {year}
	{2016})}\BibitemShut {NoStop}%
\bibitem [{\citenamefont {Alicea}\ \emph {et~al.}(2011)\citenamefont {Alicea},
	\citenamefont {Oreg}, \citenamefont {Refael}, \citenamefont {von Oppen},\
	and\ \citenamefont {Fisher}}]{Alicea2011:NP}%
\BibitemOpen
\bibfield  {author} {\bibinfo {author} {\bibfnamefont {J.}~\bibnamefont
		{Alicea}}, \bibinfo {author} {\bibfnamefont {Y.}~\bibnamefont {Oreg}},
	\bibinfo {author} {\bibfnamefont {G.}~\bibnamefont {Refael}}, \bibinfo
	{author} {\bibfnamefont {F.}~\bibnamefont {von Oppen}}, \ and\ \bibinfo
	{author} {\bibfnamefont {M.~P.~A.}\ \bibnamefont {Fisher}},\ }\href {\doibase
	10.1038/nphys1915} {\bibfield  {journal} {\bibinfo  {journal} {Nature Phys.}\
	}\textbf {\bibinfo {volume} {7}},\ \bibinfo {pages} {412} (\bibinfo {year}
	{2011})}\BibitemShut {NoStop}%
\bibitem [{\citenamefont {Aasen}\ \emph {et~al.}(2016)\citenamefont {Aasen},
	\citenamefont {Hell}, \citenamefont {Mishmash}, \citenamefont {Higginbotham},
	\citenamefont {Danon}, \citenamefont {Leijnse}, \citenamefont {Jespersen},
	\citenamefont {Folk}, \citenamefont {Marcus}, \citenamefont {Flensberg},\
	and\ \citenamefont {Alicea}}]{Aasen2016:PRX}%
\BibitemOpen
\bibfield  {author} {\bibinfo {author} {\bibfnamefont {D.}~\bibnamefont
		{Aasen}}, \bibinfo {author} {\bibfnamefont {M.}~\bibnamefont {Hell}},
	\bibinfo {author} {\bibfnamefont {R.~V.}\ \bibnamefont {Mishmash}}, \bibinfo
	{author} {\bibfnamefont {A.}~\bibnamefont {Higginbotham}}, \bibinfo {author}
	{\bibfnamefont {J.}~\bibnamefont {Danon}}, \bibinfo {author} {\bibfnamefont
		{M.}~\bibnamefont {Leijnse}}, \bibinfo {author} {\bibfnamefont {T.~S.}\
		\bibnamefont {Jespersen}}, \bibinfo {author} {\bibfnamefont {J.~A.}\
		\bibnamefont {Folk}}, \bibinfo {author} {\bibfnamefont {C.~M.}\ \bibnamefont
		{Marcus}}, \bibinfo {author} {\bibfnamefont {K.}~\bibnamefont {Flensberg}}, \
	and\ \bibinfo {author} {\bibfnamefont {J.}~\bibnamefont {Alicea}},\ }\href
{\doibase 10.1103/PhysRevX.6.031016} {\bibfield  {journal} {\bibinfo
		{journal} {Phys. Rev. X}\ }\textbf {\bibinfo {volume} {6}},\ \bibinfo {pages}
	{031016} (\bibinfo {year} {2016})}\BibitemShut {NoStop}%
\bibitem [{\citenamefont {Potter}\ and\ \citenamefont
	{Lee}(2010)}]{Potter2010:PRL}%
\BibitemOpen
\bibfield  {author} {\bibinfo {author} {\bibfnamefont {A.~C.}\ \bibnamefont
		{Potter}}\ and\ \bibinfo {author} {\bibfnamefont {P.~A.}\ \bibnamefont
		{Lee}},\ }\href {\doibase 10.1103/PhysRevLett.105.227003} {\bibfield
	{journal} {\bibinfo  {journal} {Phys. Rev. Lett.}\ }\textbf {\bibinfo
		{volume} {105}},\ \bibinfo {pages} {227003} (\bibinfo {year}
	{2010})}\BibitemShut {NoStop}%
\bibitem [{\citenamefont {Sedlmayr}\ and\ \citenamefont
	{Bena}(2015)}]{Sedlmayr2015:PRB}%
\BibitemOpen
\bibfield  {author} {\bibinfo {author} {\bibfnamefont {N.}~\bibnamefont
		{Sedlmayr}}\ and\ \bibinfo {author} {\bibfnamefont {C.}~\bibnamefont
		{Bena}},\ }\href {\doibase 10.1103/PhysRevB.92.115115} {\bibfield  {journal}
	{\bibinfo  {journal} {Phys. Rev. B}\ }\textbf {\bibinfo {volume} {92}},\
	\bibinfo {pages} {115115} (\bibinfo {year} {2015})}\BibitemShut {NoStop}%
\bibitem [{\citenamefont {Klinovaja}\ \emph {et~al.}(2012)\citenamefont
	{Klinovaja}, \citenamefont {Stano},\ and\ \citenamefont
	{Loss}}]{Klinovaja2012:PRL}%
\BibitemOpen
\bibfield  {author} {\bibinfo {author} {\bibfnamefont {J.}~\bibnamefont
		{Klinovaja}}, \bibinfo {author} {\bibfnamefont {P.}~\bibnamefont {Stano}}, \
	and\ \bibinfo {author} {\bibfnamefont {D.}~\bibnamefont {Loss}},\ }\href
{\doibase 10.1103/PhysRevLett.109.236801} {\bibfield  {journal} {\bibinfo
		{journal} {Phys. Rev. Lett.}\ }\textbf {\bibinfo {volume} {109}},\ \bibinfo
	{pages} {236801} (\bibinfo {year} {2012})}\BibitemShut {NoStop}%
\bibitem [{\citenamefont {Kjaergaard}\ \emph {et~al.}(2012)\citenamefont
	{Kjaergaard}, \citenamefont {W\"olms},\ and\ \citenamefont
	{Flensberg}}]{Kjaergaard2012:PRB}%
\BibitemOpen
\bibfield  {author} {\bibinfo {author} {\bibfnamefont {M.}~\bibnamefont
		{Kjaergaard}}, \bibinfo {author} {\bibfnamefont {K.}~\bibnamefont {W\"olms}},
	\ and\ \bibinfo {author} {\bibfnamefont {K.}~\bibnamefont {Flensberg}},\
}\href {\doibase 10.1103/PhysRevB.85.020503} {\bibfield  {journal} {\bibinfo
	{journal} {Phys. Rev. B}\ }\textbf {\bibinfo {volume} {85}},\ \bibinfo
{pages} {020503} (\bibinfo {year} {2012})}\BibitemShut {NoStop}%
\bibitem [{\citenamefont {Pientka}\ \emph {et~al.}(2013)\citenamefont
	{Pientka}, \citenamefont {Glazman},\ and\ \citenamefont {von
		Oppen}}]{Pientka2013:PRB}%
\BibitemOpen
\bibfield  {author} {\bibinfo {author} {\bibfnamefont {F.}~\bibnamefont
		{Pientka}}, \bibinfo {author} {\bibfnamefont {L.~I.}\ \bibnamefont
		{Glazman}}, \ and\ \bibinfo {author} {\bibfnamefont {F.}~\bibnamefont {von
			Oppen}},\ }\href {\doibase 10.1103/PhysRevB.88.155420} {\bibfield  {journal}
	{\bibinfo  {journal} {Phys. Rev. B}\ }\textbf {\bibinfo {volume} {88}},\
	\bibinfo {pages} {155420} (\bibinfo {year} {2013})}\BibitemShut {NoStop}%
\bibitem [{\citenamefont {Fatin}\ \emph {et~al.}(2016)\citenamefont {Fatin},
	\citenamefont {Matos-Abiague}, \citenamefont {Scharf},\ and\ \citenamefont
	{\ifmmode \check{Z}\else \v{Z}\fi{}uti\ifmmode~\acute{c}\else
		\'{c}\fi{}}}]{Fatin2016:PRL}%
\BibitemOpen
\bibfield  {author} {\bibinfo {author} {\bibfnamefont {G.~L.}\ \bibnamefont
		{Fatin}}, \bibinfo {author} {\bibfnamefont {A.}~\bibnamefont
		{Matos-Abiague}}, \bibinfo {author} {\bibfnamefont {B.}~\bibnamefont
		{Scharf}}, \ and\ \bibinfo {author} {\bibfnamefont {I.}~\bibnamefont
		{\ifmmode \check{Z}\else \v{Z}\fi{}uti\ifmmode~\acute{c}\else \'{c}\fi{}}},\
}\href {\doibase 10.1103/PhysRevLett.117.077002} {\bibfield  {journal}
{\bibinfo  {journal} {Phys. Rev. Lett.}\ }\textbf {\bibinfo {volume} {117}},\
\bibinfo {pages} {077002} (\bibinfo {year} {2016})}\BibitemShut {NoStop}%
\bibitem [{\citenamefont {Shabani}\ \emph {et~al.}(2016)\citenamefont
	{Shabani}, \citenamefont {Kjaergaard}, \citenamefont {Suominen},
	\citenamefont {Kim}, \citenamefont {Nichele}, \citenamefont {Pakrouski},
	\citenamefont {Stankevic}, \citenamefont {Lutchyn}, \citenamefont
	{Krogstrup}, \citenamefont {Feidenhans'l}, \citenamefont {Kraemer},
	\citenamefont {Nayak}, \citenamefont {Troyer}, \citenamefont {Marcus},\ and\
	\citenamefont {Palmstr\o{}m}}]{Shabani2016:PRB}%
\BibitemOpen
\bibfield  {author} {\bibinfo {author} {\bibfnamefont {J.}~\bibnamefont
		{Shabani}}, \bibinfo {author} {\bibfnamefont {M.}~\bibnamefont {Kjaergaard}},
	\bibinfo {author} {\bibfnamefont {H.~J.}\ \bibnamefont {Suominen}}, \bibinfo
	{author} {\bibfnamefont {Y.}~\bibnamefont {Kim}}, \bibinfo {author}
	{\bibfnamefont {F.}~\bibnamefont {Nichele}}, \bibinfo {author} {\bibfnamefont
		{K.}~\bibnamefont {Pakrouski}}, \bibinfo {author} {\bibfnamefont
		{T.}~\bibnamefont {Stankevic}}, \bibinfo {author} {\bibfnamefont {R.~M.}\
		\bibnamefont {Lutchyn}}, \bibinfo {author} {\bibfnamefont {P.}~\bibnamefont
		{Krogstrup}}, \bibinfo {author} {\bibfnamefont {R.}~\bibnamefont
		{Feidenhans'l}}, \bibinfo {author} {\bibfnamefont {S.}~\bibnamefont
		{Kraemer}}, \bibinfo {author} {\bibfnamefont {C.}~\bibnamefont {Nayak}},
	\bibinfo {author} {\bibfnamefont {M.}~\bibnamefont {Troyer}}, \bibinfo
	{author} {\bibfnamefont {C.~M.}\ \bibnamefont {Marcus}}, \ and\ \bibinfo
	{author} {\bibfnamefont {C.~J.}\ \bibnamefont {Palmstr\o{}m}},\ }\href
{\doibase 10.1103/PhysRevB.93.155402} {\bibfield  {journal} {\bibinfo
		{journal} {Phys. Rev. B}\ }\textbf {\bibinfo {volume} {93}},\ \bibinfo
	{pages} {155402} (\bibinfo {year} {2016})}\BibitemShut {NoStop}%
\bibitem [{\citenamefont {{Suominen}}\ \emph {et~al.}(2017)\citenamefont
	{{Suominen}}, \citenamefont {{Kjaergaard}}, \citenamefont {{Hamilton}},
	\citenamefont {{Shabani}}, \citenamefont {{Palmstr{\o}m}}, \citenamefont
	{{Marcus}},\ and\ \citenamefont {{Nichele}}}]{Suominen2017:P}%
\BibitemOpen
\bibfield  {author} {\bibinfo {author} {\bibfnamefont {H.~J.}\ \bibnamefont
		{{Suominen}}}, \bibinfo {author} {\bibfnamefont {M.}~\bibnamefont
		{{Kjaergaard}}}, \bibinfo {author} {\bibfnamefont {A.~R.}\ \bibnamefont
		{{Hamilton}}}, \bibinfo {author} {\bibfnamefont {J.}~\bibnamefont
		{{Shabani}}}, \bibinfo {author} {\bibfnamefont {C.~J.}\ \bibnamefont
		{{Palmstr{\o}m}}}, \bibinfo {author} {\bibfnamefont {C.~M.}\ \bibnamefont
		{{Marcus}}}, \ and\ \bibinfo {author} {\bibfnamefont {F.}~\bibnamefont
		{{Nichele}}},\ }\href@noop {} {\bibfield  {journal} {\bibinfo  {journal}
		{arXiv:1703.03699}\ } (\bibinfo {year} {2017})}\BibitemShut {NoStop}%
\bibitem [{\citenamefont {Kim}\ \emph {et~al.}(2015)\citenamefont {Kim},
	\citenamefont {Tewari},\ and\ \citenamefont {Tserkovnyak}}]{Kim2015:PRB}%
\BibitemOpen
\bibfield  {author} {\bibinfo {author} {\bibfnamefont {S.~K.}\ \bibnamefont
		{Kim}}, \bibinfo {author} {\bibfnamefont {S.}~\bibnamefont {Tewari}}, \ and\
	\bibinfo {author} {\bibfnamefont {Y.}~\bibnamefont {Tserkovnyak}},\ }\href
{\doibase 10.1103/PhysRevB.92.020412} {\bibfield  {journal} {\bibinfo
		{journal} {Phys. Rev. B}\ }\textbf {\bibinfo {volume} {92}},\ \bibinfo
	{pages} {020412} (\bibinfo {year} {2015})}\BibitemShut {NoStop}%
\bibitem [{\citenamefont {Sau}\ \emph {et~al.}(2011)\citenamefont {Sau},
	\citenamefont {Clarke},\ and\ \citenamefont {Tewari}}]{Sau2011:PRB}%
\BibitemOpen
\bibfield  {author} {\bibinfo {author} {\bibfnamefont {J.~D.}\ \bibnamefont
		{Sau}}, \bibinfo {author} {\bibfnamefont {D.~J.}\ \bibnamefont {Clarke}}, \
	and\ \bibinfo {author} {\bibfnamefont {S.}~\bibnamefont {Tewari}},\ }\href
{\doibase 10.1103/PhysRevB.84.094505} {\bibfield  {journal} {\bibinfo
		{journal} {Phys. Rev. B}\ }\textbf {\bibinfo {volume} {84}},\ \bibinfo
	{pages} {094505} (\bibinfo {year} {2011})}\BibitemShut {NoStop}%
\bibitem [{\citenamefont {Clarke}\ \emph {et~al.}(2011)\citenamefont {Clarke},
	\citenamefont {Sau},\ and\ \citenamefont {Tewari}}]{Clarke2011:PRB}%
\BibitemOpen
\bibfield  {author} {\bibinfo {author} {\bibfnamefont {D.~J.}\ \bibnamefont
		{Clarke}}, \bibinfo {author} {\bibfnamefont {J.~D.}\ \bibnamefont {Sau}}, \
	and\ \bibinfo {author} {\bibfnamefont {S.}~\bibnamefont {Tewari}},\ }\href
{\doibase 10.1103/PhysRevB.84.035120} {\bibfield  {journal} {\bibinfo
		{journal} {Phys. Rev. B}\ }\textbf {\bibinfo {volume} {84}},\ \bibinfo
	{pages} {035120} (\bibinfo {year} {2011})}\BibitemShut {NoStop}%
\bibitem [{\citenamefont {Halperin}\ \emph {et~al.}(2012)\citenamefont
	{Halperin}, \citenamefont {Oreg}, \citenamefont {Stern}, \citenamefont
	{Refael}, \citenamefont {Alicea},\ and\ \citenamefont {von
		Oppen}}]{Halperin2012:PRB}%
\BibitemOpen
\bibfield  {author} {\bibinfo {author} {\bibfnamefont {B.~I.}\ \bibnamefont
		{Halperin}}, \bibinfo {author} {\bibfnamefont {Y.}~\bibnamefont {Oreg}},
	\bibinfo {author} {\bibfnamefont {A.}~\bibnamefont {Stern}}, \bibinfo
	{author} {\bibfnamefont {G.}~\bibnamefont {Refael}}, \bibinfo {author}
	{\bibfnamefont {J.}~\bibnamefont {Alicea}}, \ and\ \bibinfo {author}
	{\bibfnamefont {F.}~\bibnamefont {von Oppen}},\ }\href {\doibase
	10.1103/PhysRevB.85.144501} {\bibfield  {journal} {\bibinfo  {journal} {Phys.
			Rev. B}\ }\textbf {\bibinfo {volume} {85}},\ \bibinfo {pages} {144501}
	(\bibinfo {year} {2012})}\BibitemShut {NoStop}%
\bibitem [{\citenamefont {Klinovaja}\ and\ \citenamefont
	{Loss}(2013)}]{Klinovaja2013:PRX}%
\BibitemOpen
\bibfield  {author} {\bibinfo {author} {\bibfnamefont {J.}~\bibnamefont
		{Klinovaja}}\ and\ \bibinfo {author} {\bibfnamefont {D.}~\bibnamefont
		{Loss}},\ }\href {\doibase 10.1103/PhysRevX.3.011008} {\bibfield  {journal}
	{\bibinfo  {journal} {Phys. Rev. X}\ }\textbf {\bibinfo {volume} {3}},\
	\bibinfo {pages} {011008} (\bibinfo {year} {2013})}\BibitemShut {NoStop}%
\bibitem [{\citenamefont {Korenman}\ \emph {et~al.}(1977)\citenamefont
	{Korenman}, \citenamefont {Murray},\ and\ \citenamefont
	{Prange}}]{Korenman1977:PRB}%
\BibitemOpen
\bibfield  {author} {\bibinfo {author} {\bibfnamefont {V.}~\bibnamefont
		{Korenman}}, \bibinfo {author} {\bibfnamefont {J.~L.}\ \bibnamefont
		{Murray}}, \ and\ \bibinfo {author} {\bibfnamefont {R.~E.}\ \bibnamefont
		{Prange}},\ }\href@noop {} {\bibfield  {journal} {\bibinfo  {journal} {Phys.
			Rev. B}\ }\textbf {\bibinfo {volume} {16}},\ \bibinfo {pages} {4032}
	(\bibinfo {year} {1977})}\BibitemShut {NoStop}%
\bibitem [{\citenamefont {Tatara}\ and\ \citenamefont
	{Fukuyama}(1997)}]{Tatara1997:PRL}%
\BibitemOpen
\bibfield  {author} {\bibinfo {author} {\bibfnamefont {G.}~\bibnamefont
		{Tatara}}\ and\ \bibinfo {author} {\bibfnamefont {H.}~\bibnamefont
		{Fukuyama}},\ }\href@noop {} {\bibfield  {journal} {\bibinfo  {journal}
		{Phys. Rev. Lett.}\ }\textbf {\bibinfo {volume} {78}},\ \bibinfo {pages}
	{3773} (\bibinfo {year} {1997})}\BibitemShut {NoStop}%
\bibitem [{\citenamefont {Bruno}\ \emph {et~al.}(2004)\citenamefont {Bruno},
	\citenamefont {Dugaev},\ and\ \citenamefont {Taillefumier}}]{Bruno2004:PRL}%
\BibitemOpen
\bibfield  {author} {\bibinfo {author} {\bibfnamefont {P.}~\bibnamefont
		{Bruno}}, \bibinfo {author} {\bibfnamefont {V.~K.}\ \bibnamefont {Dugaev}}, \
	and\ \bibinfo {author} {\bibfnamefont {M.}~\bibnamefont {Taillefumier}},\
}\href {\doibase 10.1103/PhysRevLett.93.096806} {\bibfield  {journal}
{\bibinfo  {journal} {Phys. Rev. Lett.}\ }\textbf {\bibinfo {volume} {93}},\
\bibinfo {pages} {096806} (\bibinfo {year} {2004})}\BibitemShut {NoStop}%
\bibitem [{\citenamefont {Jia}\ and\ \citenamefont
	{Berakdar}(2010)}]{Jia2010:PRB}%
\BibitemOpen
\bibfield  {author} {\bibinfo {author} {\bibfnamefont {C.}~\bibnamefont
		{Jia}}\ and\ \bibinfo {author} {\bibfnamefont {J.}~\bibnamefont {Berakdar}},\
}\href {\doibase 10.1103/PhysRevB.81.052406} {\bibfield  {journal} {\bibinfo
	{journal} {Phys. Rev. B}\ }\textbf {\bibinfo {volume} {81}},\ \bibinfo
{pages} {052406} (\bibinfo {year} {2010})}\BibitemShut {NoStop}%
\bibitem [{\citenamefont {Braunecker}\ \emph {et~al.}(2010)\citenamefont
	{Braunecker}, \citenamefont {Japaridze}, \citenamefont {Klinovaja},\ and\
	\citenamefont {Loss}}]{Braunecker2010:PRB}%
\BibitemOpen
\bibfield  {author} {\bibinfo {author} {\bibfnamefont {B.}~\bibnamefont
		{Braunecker}}, \bibinfo {author} {\bibfnamefont {G.~I.}\ \bibnamefont
		{Japaridze}}, \bibinfo {author} {\bibfnamefont {J.}~\bibnamefont
		{Klinovaja}}, \ and\ \bibinfo {author} {\bibfnamefont {D.}~\bibnamefont
		{Loss}},\ }\href@noop {} {\bibfield  {journal} {\bibinfo  {journal} {Phys.
			Rev. B}\ }\textbf {\bibinfo {volume} {82}},\ \bibinfo {pages} {045127}
	(\bibinfo {year} {2010})}\BibitemShut {NoStop}%
\bibitem [{\citenamefont {Klinovaja}\ \emph {et~al.}(2013)\citenamefont
	{Klinovaja}, \citenamefont {Stano}, \citenamefont {Yazdani},\ and\
	\citenamefont {Loss}}]{Klinovaja2013:PRL}%
\BibitemOpen
\bibfield  {author} {\bibinfo {author} {\bibfnamefont {J.}~\bibnamefont
		{Klinovaja}}, \bibinfo {author} {\bibfnamefont {P.}~\bibnamefont {Stano}},
	\bibinfo {author} {\bibfnamefont {A.}~\bibnamefont {Yazdani}}, \ and\
	\bibinfo {author} {\bibfnamefont {D.}~\bibnamefont {Loss}},\ }\href {\doibase
	10.1103/PhysRevLett.111.186805} {\bibfield  {journal} {\bibinfo  {journal}
		{Phys. Rev. Lett.}\ }\textbf {\bibinfo {volume} {111}},\ \bibinfo {pages}
	{186805} (\bibinfo {year} {2013})}\BibitemShut {NoStop}%
\bibitem [{\citenamefont {Vazifeh}\ and\ \citenamefont
	{Franz}(2013)}]{Vazifeh2013:PRL}%
\BibitemOpen
\bibfield  {author} {\bibinfo {author} {\bibfnamefont {M.~M.}\ \bibnamefont
		{Vazifeh}}\ and\ \bibinfo {author} {\bibfnamefont {M.}~\bibnamefont
		{Franz}},\ }\href@noop {} {\bibfield  {journal} {\bibinfo  {journal} {Phys.
			Rev. Lett.}\ }\textbf {\bibinfo {volume} {111}},\ \bibinfo {pages} {206802}
	(\bibinfo {year} {2013})}\BibitemShut {NoStop}%
\bibitem [{\citenamefont {Marra}\ and\ \citenamefont
	{Cuoco}(2017)}]{Marra2017:PRB}%
\BibitemOpen
\bibfield  {author} {\bibinfo {author} {\bibfnamefont {P.}~\bibnamefont
		{Marra}}\ and\ \bibinfo {author} {\bibfnamefont {M.}~\bibnamefont {Cuoco}},\
}\href {\doibase 10.1103/PhysRevB.95.140504} {\bibfield  {journal} {\bibinfo
	{journal} {Phys. Rev. B}\ }\textbf {\bibinfo {volume} {95}},\ \bibinfo
{pages} {140504} (\bibinfo {year} {2017})}\BibitemShut {NoStop}%
\bibitem [{\citenamefont {Nakosai}\ \emph {et~al.}(2013)\citenamefont
	{Nakosai}, \citenamefont {Tanaka},\ and\ \citenamefont
	{Nagaosa}}]{Nakosai2013:PRB}%
\BibitemOpen
\bibfield  {author} {\bibinfo {author} {\bibfnamefont {S.}~\bibnamefont
		{Nakosai}}, \bibinfo {author} {\bibfnamefont {Y.}~\bibnamefont {Tanaka}}, \
	and\ \bibinfo {author} {\bibfnamefont {N.}~\bibnamefont {Nagaosa}},\ }\href
{\doibase 10.1103/PhysRevB.88.180503} {\bibfield  {journal} {\bibinfo
		{journal} {Phys. Rev. B}\ }\textbf {\bibinfo {volume} {88}},\ \bibinfo
	{pages} {180503} (\bibinfo {year} {2013})}\BibitemShut {NoStop}%
\bibitem [{\citenamefont {Kim}\ \emph {et~al.}(2014)\citenamefont {Kim},
	\citenamefont {Cheng}, \citenamefont {Bauer}, \citenamefont {Lutchyn},\ and\
	\citenamefont {Das~Sarma}}]{Kim2014:PRB}%
\BibitemOpen
\bibfield  {author} {\bibinfo {author} {\bibfnamefont {Y.}~\bibnamefont
		{Kim}}, \bibinfo {author} {\bibfnamefont {M.}~\bibnamefont {Cheng}}, \bibinfo
	{author} {\bibfnamefont {B.}~\bibnamefont {Bauer}}, \bibinfo {author}
	{\bibfnamefont {R.~M.}\ \bibnamefont {Lutchyn}}, \ and\ \bibinfo {author}
	{\bibfnamefont {S.}~\bibnamefont {Das~Sarma}},\ }\href {\doibase
	10.1103/PhysRevB.90.060401} {\bibfield  {journal} {\bibinfo  {journal} {Phys.
			Rev. B}\ }\textbf {\bibinfo {volume} {90}},\ \bibinfo {pages} {060401}
	(\bibinfo {year} {2014})}\BibitemShut {NoStop}%
\bibitem [{\citenamefont {Schecter}\ \emph {et~al.}(2016)\citenamefont
	{Schecter}, \citenamefont {Flensberg}, \citenamefont {Christensen},
	\citenamefont {Andersen},\ and\ \citenamefont {Paaske}}]{Schecter2016:PRB}%
\BibitemOpen
\bibfield  {author} {\bibinfo {author} {\bibfnamefont {M.}~\bibnamefont
		{Schecter}}, \bibinfo {author} {\bibfnamefont {K.}~\bibnamefont {Flensberg}},
	\bibinfo {author} {\bibfnamefont {M.~H.}\ \bibnamefont {Christensen}},
	\bibinfo {author} {\bibfnamefont {B.~M.}\ \bibnamefont {Andersen}}, \ and\
	\bibinfo {author} {\bibfnamefont {J.}~\bibnamefont {Paaske}},\ }\href
{\doibase 10.1103/PhysRevB.93.140503} {\bibfield  {journal} {\bibinfo
		{journal} {Phys. Rev. B}\ }\textbf {\bibinfo {volume} {93}},\ \bibinfo
	{pages} {140503} (\bibinfo {year} {2016})}\BibitemShut {NoStop}%
\bibitem [{\citenamefont {Yang}\ \emph {et~al.}(2016)\citenamefont {Yang},
	\citenamefont {Stano}, \citenamefont {Klinovaja},\ and\ \citenamefont
	{Loss}}]{Yang2016:PRB}%
\BibitemOpen
\bibfield  {author} {\bibinfo {author} {\bibfnamefont {G.}~\bibnamefont
		{Yang}}, \bibinfo {author} {\bibfnamefont {P.}~\bibnamefont {Stano}},
	\bibinfo {author} {\bibfnamefont {J.}~\bibnamefont {Klinovaja}}, \ and\
	\bibinfo {author} {\bibfnamefont {D.}~\bibnamefont {Loss}},\ }\href {\doibase
	10.1103/PhysRevB.93.224505} {\bibfield  {journal} {\bibinfo  {journal} {Phys.
			Rev. B}\ }\textbf {\bibinfo {volume} {93}},\ \bibinfo {pages} {224505}
	(\bibinfo {year} {2016})}\BibitemShut {NoStop}%
\bibitem [{\citenamefont {Xu}\ \emph {et~al.}(2015)\citenamefont {Xu},
	\citenamefont {Wang}, \citenamefont {Liu}, \citenamefont {Ge}, \citenamefont
	{Yang}, \citenamefont {Liu}, \citenamefont {Xu}, \citenamefont {Guan},
	\citenamefont {Gao}, \citenamefont {Qian}, \citenamefont {Liu}, \citenamefont
	{Wang}, \citenamefont {Zhang}, \citenamefont {Xue},\ and\ \citenamefont
	{Jia}}]{Xu2015:PRL}%
\BibitemOpen
\bibfield  {author} {\bibinfo {author} {\bibfnamefont {J.-P.}\ \bibnamefont
		{Xu}}, \bibinfo {author} {\bibfnamefont {M.-X.}\ \bibnamefont {Wang}},
	\bibinfo {author} {\bibfnamefont {Z.~L.}\ \bibnamefont {Liu}}, \bibinfo
	{author} {\bibfnamefont {J.-F.}\ \bibnamefont {Ge}}, \bibinfo {author}
	{\bibfnamefont {X.}~\bibnamefont {Yang}}, \bibinfo {author} {\bibfnamefont
		{C.}~\bibnamefont {Liu}}, \bibinfo {author} {\bibfnamefont {Z.~A.}\
		\bibnamefont {Xu}}, \bibinfo {author} {\bibfnamefont {D.}~\bibnamefont
		{Guan}}, \bibinfo {author} {\bibfnamefont {C.~L.}\ \bibnamefont {Gao}},
	\bibinfo {author} {\bibfnamefont {D.}~\bibnamefont {Qian}}, \bibinfo {author}
	{\bibfnamefont {Y.}~\bibnamefont {Liu}}, \bibinfo {author} {\bibfnamefont
		{Q.-H.}\ \bibnamefont {Wang}}, \bibinfo {author} {\bibfnamefont {F.-C.}\
		\bibnamefont {Zhang}}, \bibinfo {author} {\bibfnamefont {Q.-K.}\ \bibnamefont
		{Xue}}, \ and\ \bibinfo {author} {\bibfnamefont {J.-F.}\ \bibnamefont
		{Jia}},\ }\href {\doibase 10.1103/PhysRevLett.114.017001} {\bibfield
	{journal} {\bibinfo  {journal} {Phys. Rev. Lett.}\ }\textbf {\bibinfo
		{volume} {114}},\ \bibinfo {pages} {017001} (\bibinfo {year}
	{2015})}\BibitemShut {NoStop}%
\bibitem [{\citenamefont {Sun}\ \emph {et~al.}(2016)\citenamefont {Sun},
	\citenamefont {Zhang}, \citenamefont {Hu}, \citenamefont {Li}, \citenamefont
	{Wang}, \citenamefont {Ma}, \citenamefont {Xu}, \citenamefont {Gao},
	\citenamefont {Guan}, \citenamefont {Li}, \citenamefont {Liu}, \citenamefont
	{Qian}, \citenamefont {Zhou}, \citenamefont {Fu}, \citenamefont {Li},
	\citenamefont {Zhang},\ and\ \citenamefont {Jia}}]{Sun2016:PRL}%
\BibitemOpen
\bibfield  {author} {\bibinfo {author} {\bibfnamefont {H.-H.}\ \bibnamefont
		{Sun}}, \bibinfo {author} {\bibfnamefont {K.-W.}\ \bibnamefont {Zhang}},
	\bibinfo {author} {\bibfnamefont {L.-H.}\ \bibnamefont {Hu}}, \bibinfo
	{author} {\bibfnamefont {C.}~\bibnamefont {Li}}, \bibinfo {author}
	{\bibfnamefont {G.-Y.}\ \bibnamefont {Wang}}, \bibinfo {author}
	{\bibfnamefont {H.-Y.}\ \bibnamefont {Ma}}, \bibinfo {author} {\bibfnamefont
		{Z.-A.}\ \bibnamefont {Xu}}, \bibinfo {author} {\bibfnamefont {C.-L.}\
		\bibnamefont {Gao}}, \bibinfo {author} {\bibfnamefont {D.-D.}\ \bibnamefont
		{Guan}}, \bibinfo {author} {\bibfnamefont {Y.-Y.}\ \bibnamefont {Li}},
	\bibinfo {author} {\bibfnamefont {C.}~\bibnamefont {Liu}}, \bibinfo {author}
	{\bibfnamefont {D.}~\bibnamefont {Qian}}, \bibinfo {author} {\bibfnamefont
		{Y.}~\bibnamefont {Zhou}}, \bibinfo {author} {\bibfnamefont {L.}~\bibnamefont
		{Fu}}, \bibinfo {author} {\bibfnamefont {S.-C.}\ \bibnamefont {Li}}, \bibinfo
	{author} {\bibfnamefont {F.-C.}\ \bibnamefont {Zhang}}, \ and\ \bibinfo
	{author} {\bibfnamefont {J.-F.}\ \bibnamefont {Jia}},\ }\href {\doibase
	10.1103/PhysRevLett.116.257003} {\bibfield  {journal} {\bibinfo  {journal}
		{Phys. Rev. Lett.}\ }\textbf {\bibinfo {volume} {116}},\ \bibinfo {pages}
	{257003} (\bibinfo {year} {2016})}\BibitemShut {NoStop}%
\bibitem [{\citenamefont {Prada}\ \emph {et~al.}(2012)\citenamefont {Prada},
	\citenamefont {San-Jose},\ and\ \citenamefont {Aguado}}]{Prada2012:PRB}%
\BibitemOpen
\bibfield  {author} {\bibinfo {author} {\bibfnamefont {E.}~\bibnamefont
		{Prada}}, \bibinfo {author} {\bibfnamefont {P.}~\bibnamefont {San-Jose}}, \
	and\ \bibinfo {author} {\bibfnamefont {R.}~\bibnamefont {Aguado}},\ }\href
{\doibase 10.1103/PhysRevB.86.180503} {\bibfield  {journal} {\bibinfo
		{journal} {Phys. Rev. B}\ }\textbf {\bibinfo {volume} {86}},\ \bibinfo
	{pages} {180503} (\bibinfo {year} {2012})}\BibitemShut {NoStop}%
\bibitem [{\citenamefont {Franz}(2013)}]{Franz2013:NN}%
\BibitemOpen
\bibfield  {author} {\bibinfo {author} {\bibfnamefont {M.}~\bibnamefont
		{Franz}},\ }\href {\doibase 10.1038/nnano.2013.33} {\bibfield  {journal}
	{\bibinfo  {journal} {Nature Nanotech.}\ }\textbf {\bibinfo {volume} {8}},\
	\bibinfo {pages} {149} (\bibinfo {year} {2013})}\BibitemShut {NoStop}%
\bibitem [{\citenamefont {Scharf}\ and\ \citenamefont
	{\v{Z}uti\'c}(2015)}]{Scharf2015:PRB}%
\BibitemOpen
\bibfield  {author} {\bibinfo {author} {\bibfnamefont {B.}~\bibnamefont
		{Scharf}}\ and\ \bibinfo {author} {\bibfnamefont {I.}~\bibnamefont
		{\v{Z}uti\'c}},\ }\href@noop {} {\bibfield  {journal} {\bibinfo  {journal}
		{Phys. Rev. B}\ }\textbf {\bibinfo {volume} {91}},\ \bibinfo {pages} {144505}
	(\bibinfo {year} {2015})}\BibitemShut {NoStop}%
\bibitem [{\citenamefont {Adagideli}\ \emph {et~al.}(2014)\citenamefont
	{Adagideli}, \citenamefont {Wimmer},\ and\ \citenamefont
	{Teker}}]{Adagideli2014:PRB}%
\BibitemOpen
\bibfield  {author} {\bibinfo {author} {\bibfnamefont {I.}~\bibnamefont
		{Adagideli}}, \bibinfo {author} {\bibfnamefont {M.}~\bibnamefont {Wimmer}}, \
	and\ \bibinfo {author} {\bibfnamefont {A.}~\bibnamefont {Teker}},\ }\href
{\doibase 10.1103/PhysRevB.89.144506} {\bibfield  {journal} {\bibinfo
		{journal} {Phys. Rev. B}\ }\textbf {\bibinfo {volume} {89}},\ \bibinfo
	{pages} {144506} (\bibinfo {year} {2014})}\BibitemShut {NoStop}%
\bibitem [{\citenamefont {Svensson}\ \emph {et~al.}(2012)\citenamefont
	{Svensson}, \citenamefont {Sarney}, \citenamefont {Hier}, \citenamefont
	{Lin}, \citenamefont {Wang}, \citenamefont {Donetsky}, \citenamefont
	{Shterengas}, \citenamefont {Kipshidze},\ and\ \citenamefont
	{Belenky}}]{Svensson2012:PRB}%
\BibitemOpen
\bibfield  {author} {\bibinfo {author} {\bibfnamefont {S.~P.}\ \bibnamefont
		{Svensson}}, \bibinfo {author} {\bibfnamefont {W.~L.}\ \bibnamefont
		{Sarney}}, \bibinfo {author} {\bibfnamefont {H.}~\bibnamefont {Hier}},
	\bibinfo {author} {\bibfnamefont {Y.}~\bibnamefont {Lin}}, \bibinfo {author}
	{\bibfnamefont {D.}~\bibnamefont {Wang}}, \bibinfo {author} {\bibfnamefont
		{D.}~\bibnamefont {Donetsky}}, \bibinfo {author} {\bibfnamefont
		{L.}~\bibnamefont {Shterengas}}, \bibinfo {author} {\bibfnamefont
		{G.}~\bibnamefont {Kipshidze}}, \ and\ \bibinfo {author} {\bibfnamefont
		{G.}~\bibnamefont {Belenky}},\ }\href@noop {} {\bibfield  {journal} {\bibinfo
		{journal} {Phys. Rev. B}\ }\textbf {\bibinfo {volume} {86}},\ \bibinfo
	{pages} {245205} (\bibinfo {year} {2012})}\BibitemShut {NoStop}%
\bibitem [{\citenamefont {Tsymbal}\ and\ \citenamefont
	{\v{Z}uti\'c}(2011)}]{Tsymbal:2011}%
\BibitemOpen
\bibinfo {editor} {\bibfnamefont {E.~Y.}\ \bibnamefont {Tsymbal}}\ and\
\bibinfo {editor} {\bibfnamefont {I.}~\bibnamefont {\v{Z}uti\'c}},\ eds.,\
\href@noop {} {\emph {\bibinfo {title} {Handbook of Spin Transport and
			Magnetism}}}\ (\bibinfo  {publisher} {Chapman \& Hall/CRC, Boca Raton, FL},\
\bibinfo {year} {2011})\BibitemShut {NoStop}%
\bibitem [{\citenamefont {\ifmmode \check{Z}\else
		\v{Z}\fi{}uti\ifmmode~\acute{c}\else \'{c}\fi{}}\ \emph
	{et~al.}(2004)\citenamefont {\ifmmode \check{Z}\else
		\v{Z}\fi{}uti\ifmmode~\acute{c}\else \'{c}\fi{}}, \citenamefont {Fabian},\
	and\ \citenamefont {Das~Sarma}}]{Zutic2004:RMP}%
\BibitemOpen
\bibfield  {author} {\bibinfo {author} {\bibfnamefont {I.}~\bibnamefont
		{\ifmmode \check{Z}\else \v{Z}\fi{}uti\ifmmode~\acute{c}\else \'{c}\fi{}}},
	\bibinfo {author} {\bibfnamefont {J.}~\bibnamefont {Fabian}}, \ and\ \bibinfo
	{author} {\bibfnamefont {S.}~\bibnamefont {Das~Sarma}},\ }\href {\doibase
	10.1103/RevModPhys.76.323} {\bibfield  {journal} {\bibinfo  {journal} {Rev.
			Mod. Phys.}\ }\textbf {\bibinfo {volume} {76}},\ \bibinfo {pages} {323}
	(\bibinfo {year} {2004})}\BibitemShut {NoStop}%
\bibitem [{\citenamefont {Betthausen}\ \emph {et~al.}(2012)\citenamefont
	{Betthausen}, \citenamefont {Dollinger}, \citenamefont {Saarikoski},
	\citenamefont {Kolkovsky}, \citenamefont {Karczewski}, \citenamefont
	{Wojtowicz}, \citenamefont {Richter},\ and\ \citenamefont
	{Weiss}}]{Betthausen2012:S}%
\BibitemOpen
\bibfield  {author} {\bibinfo {author} {\bibfnamefont {C.}~\bibnamefont
		{Betthausen}}, \bibinfo {author} {\bibfnamefont {T.}~\bibnamefont
		{Dollinger}}, \bibinfo {author} {\bibfnamefont {H.}~\bibnamefont
		{Saarikoski}}, \bibinfo {author} {\bibfnamefont {V.}~\bibnamefont
		{Kolkovsky}}, \bibinfo {author} {\bibfnamefont {G.}~\bibnamefont
		{Karczewski}}, \bibinfo {author} {\bibfnamefont {T.}~\bibnamefont
		{Wojtowicz}}, \bibinfo {author} {\bibfnamefont {K.}~\bibnamefont {Richter}},
	\ and\ \bibinfo {author} {\bibfnamefont {D.}~\bibnamefont {Weiss}},\ }\href
{\doibase 10.1126/science.1221350} {\bibfield  {journal} {\bibinfo  {journal}
		{Science}\ }\textbf {\bibinfo {volume} {337}},\ \bibinfo {pages} {324}
	(\bibinfo {year} {2012})}\BibitemShut {NoStop}%
\bibitem [{\citenamefont {\v{Z}uti\'c}\ and\ \citenamefont
	{Lee}(2012)}]{Zutic2012:S}%
\BibitemOpen
\bibfield  {author} {\bibinfo {author} {\bibfnamefont {I.}~\bibnamefont
		{\v{Z}uti\'c}}\ and\ \bibinfo {author} {\bibfnamefont {J.}~\bibnamefont
		{Lee}},\ }\href@noop {} {\bibfield  {journal} {\bibinfo  {journal} {Science}\
	}\textbf {\bibinfo {volume} {337}},\ \bibinfo {pages} {307} (\bibinfo {year}
	{2012})}\BibitemShut {NoStop}%
\bibitem [{\citenamefont {Hu}(1994)}]{Hu1994:PRL}%
\BibitemOpen
\bibfield  {author} {\bibinfo {author} {\bibfnamefont {C.~R.}\ \bibnamefont
		{Hu}},\ }\href@noop {} {\bibfield  {journal} {\bibinfo  {journal} {Phys. Rev.
			Lett.}\ }\textbf {\bibinfo {volume} {72}},\ \bibinfo {pages} {1526} (\bibinfo
	{year} {1994})}\BibitemShut {NoStop}%
\bibitem [{\citenamefont {Tanaka}\ and\ \citenamefont
	{Kashiwaya}(1995)}]{Kashiwaya1995:PRL}%
\BibitemOpen
\bibfield  {author} {\bibinfo {author} {\bibfnamefont {Y.}~\bibnamefont
		{Tanaka}}\ and\ \bibinfo {author} {\bibfnamefont {S.}~\bibnamefont
		{Kashiwaya}},\ }\href@noop {} {\bibfield  {journal} {\bibinfo  {journal}
		{Phys. Rev. Lett.}\ }\textbf {\bibinfo {volume} {74}},\ \bibinfo {pages}
	{3451} (\bibinfo {year} {1995})}\BibitemShut {NoStop}%
\bibitem [{\citenamefont {Wei}\ \emph {et~al.}(1998)\citenamefont {Wei},
	\citenamefont {Yeh}, \citenamefont {Garrigus},\ and\ \citenamefont
	{Strasik}}]{Wei1998:PRL}%
\BibitemOpen
\bibfield  {author} {\bibinfo {author} {\bibfnamefont {J.~Y.~T.}\
		\bibnamefont {Wei}}, \bibinfo {author} {\bibfnamefont {N.-C.}\ \bibnamefont
		{Yeh}}, \bibinfo {author} {\bibfnamefont {D.~F.}\ \bibnamefont {Garrigus}}, \
	and\ \bibinfo {author} {\bibfnamefont {M.}~\bibnamefont {Strasik}},\
}\href@noop {} {\bibfield  {journal} {\bibinfo  {journal} {Phys. Rev. Lett.}\
}\textbf {\bibinfo {volume} {81}},\ \bibinfo {pages} {2542} (\bibinfo {year}
{1998})}\BibitemShut {NoStop}%
\bibitem [{\citenamefont {Kashiwaya}\ and\ \citenamefont
	{Tanaka}(2000)}]{Kashiwaya2000:RPP}%
\BibitemOpen
\bibfield  {author} {\bibinfo {author} {\bibfnamefont {S.}~\bibnamefont
		{Kashiwaya}}\ and\ \bibinfo {author} {\bibfnamefont {Y.}~\bibnamefont
		{Tanaka}},\ }\href@noop {} {\bibfield  {journal} {\bibinfo  {journal} {Rep.
			Prog. Phys.}\ }\textbf {\bibinfo {volume} {63}},\ \bibinfo {pages} {1641}
	(\bibinfo {year} {2000})}\BibitemShut {NoStop}%
\bibitem [{\citenamefont {{\v{Z}uti\'c}}\ and\ \citenamefont
	{Valls}(2000)}]{Zutic2000:PRB}%
\BibitemOpen
\bibfield  {author} {\bibinfo {author} {\bibfnamefont {I.}~\bibnamefont
		{{\v{Z}uti\'c}}}\ and\ \bibinfo {author} {\bibfnamefont {O.~T.}\ \bibnamefont
		{Valls}},\ }\href@noop {} {\bibfield  {journal} {\bibinfo  {journal} {Phys.
			Rev. B}\ }\textbf {\bibinfo {volume} {61}},\ \bibinfo {pages} {1555}
	(\bibinfo {year} {2000})}\BibitemShut {NoStop}%
\bibitem [{\citenamefont {Chen}\ \emph {et~al.}(2001)\citenamefont {Chen},
	\citenamefont {Biswas}, \citenamefont {\v{Z}uti\'c}, \citenamefont {Wu},
	\citenamefont {Ogale}, \citenamefont {Greene},\ and\ \citenamefont
	{Venkatesan}}]{Chen2001:PRB}%
\BibitemOpen
\bibfield  {author} {\bibinfo {author} {\bibfnamefont {Z.~Y.}\ \bibnamefont
		{Chen}}, \bibinfo {author} {\bibfnamefont {A.}~\bibnamefont {Biswas}},
	\bibinfo {author} {\bibfnamefont {I.}~\bibnamefont {\v{Z}uti\'c}}, \bibinfo
	{author} {\bibfnamefont {T.}~\bibnamefont {Wu}}, \bibinfo {author}
	{\bibfnamefont {S.~B.}\ \bibnamefont {Ogale}}, \bibinfo {author}
	{\bibfnamefont {R.~L.}\ \bibnamefont {Greene}}, \ and\ \bibinfo {author}
	{\bibfnamefont {T.}~\bibnamefont {Venkatesan}},\ }\href@noop {} {\bibfield
	{journal} {\bibinfo  {journal} {Phys. Rev. B}\ }\textbf {\bibinfo {volume}
		{63}},\ \bibinfo {pages} {212508} (\bibinfo {year} {2001})}\BibitemShut
{NoStop}%
\bibitem [{\citenamefont {Liu}\ \emph {et~al.}(2017)\citenamefont {Liu},
	\citenamefont {Sau}, \citenamefont {Stanescu},\ and\ \citenamefont
	{Sarma}}]{Liu2017:P}%
\BibitemOpen
\bibfield  {author} {\bibinfo {author} {\bibfnamefont {C.-X.}\ \bibnamefont
		{Liu}}, \bibinfo {author} {\bibfnamefont {J.~D.}\ \bibnamefont {Sau}},
	\bibinfo {author} {\bibfnamefont {T.~D.}\ \bibnamefont {Stanescu}}, \ and\
	\bibinfo {author} {\bibfnamefont {S.~D.}\ \bibnamefont {Sarma}},\ }\href@noop
{} {\bibfield  {journal} {\bibinfo  {journal} {arXiv:1705.02035}\ } (\bibinfo
	{year} {2017})}\BibitemShut {NoStop}%
\bibitem [{\citenamefont {Liu}\ \emph {et~al.}(2012)\citenamefont {Liu},
	\citenamefont {Potter}, \citenamefont {Law},\ and\ \citenamefont
	{Lee}}]{Liu2012:PRL}%
\BibitemOpen
\bibfield  {author} {\bibinfo {author} {\bibfnamefont {J.}~\bibnamefont
		{Liu}}, \bibinfo {author} {\bibfnamefont {A.~C.}\ \bibnamefont {Potter}},
	\bibinfo {author} {\bibfnamefont {K.~T.}\ \bibnamefont {Law}}, \ and\
	\bibinfo {author} {\bibfnamefont {P.~A.}\ \bibnamefont {Lee}},\ }\href
{\doibase 10.1103/PhysRevLett.109.267002} {\bibfield  {journal} {\bibinfo
		{journal} {Phys. Rev. Lett.}\ }\textbf {\bibinfo {volume} {109}},\ \bibinfo
	{pages} {267002} (\bibinfo {year} {2012})}\BibitemShut {NoStop}%
\bibitem [{\citenamefont {{Tsuei}}(2013)}]{Tsuei2013:Arxiv}%
\BibitemOpen
\bibfield  {author} {\bibinfo {author} {\bibfnamefont {C.~C.}\ \bibnamefont
		{{Tsuei}}},\ }\href@noop {} {\bibfield  {journal} {\bibinfo  {journal}
		{arXiv:1306.0652}\ } (\bibinfo {year} {2013})}\BibitemShut {NoStop}%
\bibitem [{\citenamefont {Winkler}\ \emph {et~al.}(2017)\citenamefont
	{Winkler}, \citenamefont {Varjas}, \citenamefont {Skolasinski}, \citenamefont
	{Soluyanov}, \citenamefont {Troyer},\ and\ \citenamefont
	{Wimmer}}]{Winkler2017:P}%
\BibitemOpen
\bibfield  {author} {\bibinfo {author} {\bibfnamefont {G.~W.}\ \bibnamefont
		{Winkler}}, \bibinfo {author} {\bibfnamefont {D.}~\bibnamefont {Varjas}},
	\bibinfo {author} {\bibfnamefont {R.}~\bibnamefont {Skolasinski}}, \bibinfo
	{author} {\bibfnamefont {A.~A.}\ \bibnamefont {Soluyanov}}, \bibinfo {author}
	{\bibfnamefont {M.}~\bibnamefont {Troyer}}, \ and\ \bibinfo {author}
	{\bibfnamefont {M.}~\bibnamefont {Wimmer}},\ }\href@noop {} {\bibfield
	{journal} {\bibinfo  {journal} {arXiv:1703.10091}\ } (\bibinfo {year}
	{2017})}\BibitemShut {NoStop}%
\bibitem [{\citenamefont {Kent}\ and\ \citenamefont
	{Worledge}(2015)}]{Kent2015:NN}%
\BibitemOpen
\bibfield  {author} {\bibinfo {author} {\bibfnamefont {A.~D.}\ \bibnamefont
		{Kent}}\ and\ \bibinfo {author} {\bibfnamefont {D.~C.}\ \bibnamefont
		{Worledge}},\ }\href@noop {} {\bibfield  {journal} {\bibinfo  {journal}
		{Nature Nanotech.}\ }\textbf {\bibinfo {volume} {10}},\ \bibinfo {pages}
	{187} (\bibinfo {year} {2015})}\BibitemShut {NoStop}%
\bibitem [{\citenamefont {Hahn}\ \emph {et~al.}(2016)\citenamefont {Hahn},
	\citenamefont {Wolf}, \citenamefont {Kardasz}, \citenamefont {Watts},
	\citenamefont {Pinarbasi},\ and\ \citenamefont {Kent}}]{Hahn2016:PRB}%
\BibitemOpen
\bibfield  {author} {\bibinfo {author} {\bibfnamefont {C.}~\bibnamefont
		{Hahn}}, \bibinfo {author} {\bibfnamefont {G.}~\bibnamefont {Wolf}}, \bibinfo
	{author} {\bibfnamefont {B.}~\bibnamefont {Kardasz}}, \bibinfo {author}
	{\bibfnamefont {S.}~\bibnamefont {Watts}}, \bibinfo {author} {\bibfnamefont
		{M.}~\bibnamefont {Pinarbasi}}, \ and\ \bibinfo {author} {\bibfnamefont
		{A.~D.}\ \bibnamefont {Kent}},\ }\href {\doibase 10.1103/PhysRevB.94.214432}
{\bibfield  {journal} {\bibinfo  {journal} {Phys. Rev. B}\ }\textbf {\bibinfo
		{volume} {94}},\ \bibinfo {pages} {214432} (\bibinfo {year}
	{2016})}\BibitemShut {NoStop}%
\bibitem [{\citenamefont {Nowak}\ \emph {et~al.}(2016)\citenamefont {Nowak},
	\citenamefont {Robertazzi}, \citenamefont {Sun}, \citenamefont {Hu},
	\citenamefont {Park}, \citenamefont {Lee}, \citenamefont {Annunziata},
	\citenamefont {Lauer}, \citenamefont {Kothandaraman}, \citenamefont
	{O{'}Sullivan}, \citenamefont {Trouilloud}, \citenamefont {Kim},\ and\
	\citenamefont {Worledge}}]{Nowak2016:IEEEML}%
\BibitemOpen
\bibfield  {author} {\bibinfo {author} {\bibfnamefont {J.~J.}\ \bibnamefont
		{Nowak}}, \bibinfo {author} {\bibfnamefont {R.~P.}\ \bibnamefont
		{Robertazzi}}, \bibinfo {author} {\bibfnamefont {J.~Z.}\ \bibnamefont {Sun}},
	\bibinfo {author} {\bibfnamefont {G.}~\bibnamefont {Hu}}, \bibinfo {author}
	{\bibfnamefont {J.~H.}\ \bibnamefont {Park}}, \bibinfo {author}
	{\bibfnamefont {J.}~\bibnamefont {Lee}}, \bibinfo {author} {\bibfnamefont
		{A.~J.}\ \bibnamefont {Annunziata}}, \bibinfo {author} {\bibfnamefont
		{G.~P.}\ \bibnamefont {Lauer}}, \bibinfo {author} {\bibfnamefont
		{R.}~\bibnamefont {Kothandaraman}}, \bibinfo {author} {\bibfnamefont {E.~J.}\
		\bibnamefont {O{'}Sullivan}}, \bibinfo {author} {\bibfnamefont {P.~L.}\
		\bibnamefont {Trouilloud}}, \bibinfo {author} {\bibfnamefont
		{Y.}~\bibnamefont {Kim}}, \ and\ \bibinfo {author} {\bibfnamefont {D.~C.}\
		\bibnamefont {Worledge}},\ }\href@noop {} {\bibfield  {journal} {\bibinfo
		{journal} {IEEE Magn. Lett.}\ }\textbf {\bibinfo {volume} {7}},\ \bibinfo
	{pages} {1} (\bibinfo {year} {2016})}\BibitemShut {NoStop}%
\bibitem [{\citenamefont {Ye}\ \emph {et~al.}(2014)\citenamefont {Ye},
	\citenamefont {Gopman}, \citenamefont {Rehm}, \citenamefont {Backes},
	\citenamefont {Wolf}, \citenamefont {Ohki}, \citenamefont {Kirichenko},
	\citenamefont {Vernik}, \citenamefont {Mukhanov},\ and\ \citenamefont
	{Kent}}]{Ye2014:JAP}%
\BibitemOpen
\bibfield  {author} {\bibinfo {author} {\bibfnamefont {L.}~\bibnamefont
		{Ye}}, \bibinfo {author} {\bibfnamefont {D.~B.}\ \bibnamefont {Gopman}},
	\bibinfo {author} {\bibfnamefont {L.}~\bibnamefont {Rehm}}, \bibinfo {author}
	{\bibfnamefont {D.}~\bibnamefont {Backes}}, \bibinfo {author} {\bibfnamefont
		{G.}~\bibnamefont {Wolf}}, \bibinfo {author} {\bibfnamefont {T.}~\bibnamefont
		{Ohki}}, \bibinfo {author} {\bibfnamefont {A.~F.}\ \bibnamefont
		{Kirichenko}}, \bibinfo {author} {\bibfnamefont {I.~V.}\ \bibnamefont
		{Vernik}}, \bibinfo {author} {\bibfnamefont {O.~A.}\ \bibnamefont
		{Mukhanov}}, \ and\ \bibinfo {author} {\bibfnamefont {A.~D.}\ \bibnamefont
		{Kent}},\ }\href@noop {} {\bibfield  {journal} {\bibinfo  {journal} {J. Appl.
			Phys.}\ }\textbf {\bibinfo {volume} {115}},\ \bibinfo {eid} {17C725}
	(\bibinfo {year} {2014})}\BibitemShut {NoStop}%
\end{thebibliography}
%

\end{document}